\documentclass[prd,aps,preprintnumbers,nofootinbib,superscriptaddress]{revtex4}

\usepackage{geometry}
\usepackage[titles]{tocloft}

\usepackage{graphicx}
\usepackage{float}
\usepackage{multirow}

\usepackage{subcaption}
\usepackage{amsmath,amsthm,amssymb}
\usepackage{tikz-feynman}
\usepackage{tikz}
\usepackage{tikz-3dplot}
\usepackage{slashed}

\usepackage{hyperref}
\newcommand{\A}{\mathcal{A}}
\newcommand{\B}{\mathcal{B}}

\newcommand{\OP}{\mathcal{O}}

\newcommand{\N}{\mathcal{N}}

\newcommand{\Ha}{\mathcal{H}}
\newcommand{\Co}{\mathcal{C}}

\newcommand{\simorderl}{\raisebox{-4pt}{$\, \stackrel{\textstyle <}{\sim} \,$}}

\begin{document}

\title{TMD Evolution Study of the $\cos 2 \phi$ Azimuthal Asymmetry in Unpolarized $J/\psi$ Production at EIC}

\author{Jelle Bor}
\email{j.bor@rug.nl}
\affiliation{Van Swinderen Institute for Particle Physics and Gravity, University of Groningen, Nijenborgh 4, 9747 AG Groningen, The Netherlands}
\affiliation{Universit\'e Paris-Saclay, CNRS, IJCLab, 91405 Orsay, France}

\author{Dani\"el Boer}
\email{d.boer@rug.nl}
\affiliation{Van Swinderen Institute for Particle Physics and Gravity, University of Groningen, Nijenborgh 4, 9747 AG Groningen, The Netherlands}

\begin{abstract}
Semi-inclusive $J/\psi$ production in electron-proton collisions is a promising process to study gluon transverse momentum distributions (TMDs) at the future Electron-Ion Collider. In this article, we improve on previous studies of the $\cos 2 \phi$ azimuthal asymmetry that arises from the linear polarization of gluons inside unpolarized protons by including TMD evolution. We find that in the TMD regime the asymmetry grows monotonically with increasing transverse momentum of the outgoing $J/\psi$, in contrast to tree level calculations with Gaussian TMDs. Our predictions for the asymmetry at EIC can become very large at larger $x$, $Q$, and transverse momenta, even larger than the positivity bound. This problem stems from the very small $b$ region and implies a range of validity of TMD factorization that is more restricted than usually expected. We also include an estimate of the nonperturbative uncertainty from the large $b$ region and we conclude that it is smaller than the largest source of uncertainty, which stems from the choice of Color Octet Long-Distance Matrix Elements.     
\end{abstract}

\date{\today}

\maketitle

\section{Introduction}
Transverse momentum dependent parton distributions functions (TMDs) describe the transverse momentum distribution of quarks or gluons inside protons, or hadrons in general. For small longitudinal momentum fraction $x$ the gluons dominate, but hardly anything is known yet about the gluon TMDs experimentally. Many theoretical proposals have been put forward that could potentially be used for extractions of gluon TMDs by investigating transverse momentum spectra and azimuthal asymmetries for bound and open heavy quark production, both in lepton–proton and in proton–proton collisions. This is because heavy quarks are very sensitive to the gluon content of hadrons, as they are predominantly produced from gluons and not intrinsically present in hadrons at small momentum fractions. Furthermore, some quarkonium states, like the $J/\psi$, are relatively straightforward to detect and numerous events can be collected. Therefore, quarkonium production can been considered as a main tool to extract gluon TMDs and many studies have appeared about this topic already  \cite{Pisano:2013cya,Godbole:2012bx,Godbole:2013bca,Godbole:2014tha,Zhang:2014vmh,Mukherjee:2015smo,Mukherjee:2016qxa,DAlesio:2017rzj,Lansberg:2017dzg,Bacchetta:2018ivt,Kishore:2018ugo,Scarpa:2019fol,Kishore:2022ddb}.

One aspect of gluon TMDs that is especially of interest is that noncollinear gluons inside an unpolarized proton can be linearly polarized \cite{Mulders:2000sh}. This manifest itself in scattering processes through azimuthal asymmetries as pointed out in \cite{Boer:2010zf,Pisano:2013cya,Boer:2016fqd}. These are the QCD analogues of the QED asymmetries that were recently observed in ultra-peripheral heavy ion collisions at RHIC, where linearly polarized photons from the colliding gold nuclei produce electron-positron pairs with $\cos 2\phi$ and $\cos 4\phi$ azimuthal asymmetries \cite{STAR:2019wlg}. In order to measure these effects for gluons, the future U.S.-based Electron-Ion Collider (EIC) is well suited, especially using heavy quarkonium production processes. In semi-inclusive electro-production of a heavy quarkonium state, such as a $J/\psi$ vector meson, the linear gluon polarization manifests itself through a $\cos 2\phi$ azimuthal asymmetry \cite{Mukherjee:2016qxa,Bacchetta:2018ivt,Kishore:2018ugo}. This observable is the main subject of this paper. Other promising processes are single or double quarkonium production at the LHC, where the linear gluon polarization manifests itself through $\cos 2\phi$ and $\cos 4\phi$ azimuthal asymmetries \cite{Boer:2012bt,Scarpa:2019fol}. The advantage of the EIC is that only one TMD is involved, whereas in proton-proton collisions always a convolution of two TMDs is probed.  

Semi-inclusive electro-production of a heavy quarkonium state at small transverse momentum is expected to be describable within TMD factorization, just like semi-inclusive deep inelastic scattering (SIDIS) for light hadron production and the Drell–Yan (DY) process \cite{Collins:2011zzd, Echevarria:2011epo, Echevarria:2012js}. This is thanks to the presence of a large energy scale, the photon virtuality $Q$, which allows to factorize the cross section description in a perturbative short-distance part that can be expanded in orders of the strong coupling constant $\alpha_s$ and a nonperturbative long-distance part that is expressed in terms of TMDs. TMDs have to be modelled, calculated using lattice QCD, or extracted from experimental data. The TMD factorization description allows to incorporate the scale evolution, commonly referred to as TMD evolution. This in turn allows to improve on predictions and subsequent extractions. 

In this article, we include TMD evolution effects at leading order in $\alpha_{s}$ in a similar way as discussed in \cite{Boer:2014tka,Bacchetta:2018ivt}. In this way, we obtain more realistic estimates for the transverse momentum spectrum and the associated azimuthal asymmetry for semi-inclusive $J/\psi$ production in unpolarized electron-proton collisions for EIC kinematics. The results depend on the perturbative Sudakov factor, which has not yet been derived for this specific process. The one of \cite{Boer:2020bbd} obtained from a matching calculation seems applicable only to light hadron production, as it has been demonstrated for $pp \to \eta_c X$ \cite{Sun:2012vc} and for open heavy quark pair production in $ep$ collisions \cite{Zhu:2013yxa} that there are no double logarithms associated to heavy quark production. Here we will implement the result of \cite{Sun:2012vc}, which leads to considerably larger asymmetry values than when following \cite{Boer:2020bbd}. There will also be considerable dependence on how the region of very small $b$ ($\sim 1/Q$) is treated, indicating the need for matching to the collinear expression earlier than usually expected, i.e.\ at  transverse momentum values below $Q/2$. All this will be addressed in detail.   

In this study there is significant uncertainty from the nonperturbative contributions. Although no model for the gluon TMDs will be assumed, there is uncertainty from the nonperturbative Sudakov factor in the TMD evolution expressions, as well as from the nonperturbative formation of the bound quarkonium state from a produced heavy quark pair. The latter is usually described within nonrelativistic QCD (NRQCD) \cite{Bodwin:1994jh}. The NRQCD framework involves another factorization: a separation of the perturbative short-distance contributions (expanded in $\alpha_s$) from the nonpertubative Long-Distance Matrix Elements (LDMEs). The relative importance of the various LDMEs is estimated by means of the heavy quark-antiquark relative velocity $v$ in the bound-state rest frame in the limit $v \ll 1$, by the velocity scaling rules. Typically, charmonium states have $v^{2} \sim 0.3$, whereas bottomonium states have $v^{2} \sim 0.1$, which are considered small enough to expand in.

In TMD factorization the formation of the bound quarkonium state from a produced heavy quark pair has to be incorporated by introducing Shape Functions (SFs) \cite{Echevarria:2019ynx, Fleming:2019pzj} that can be viewed as describing the transverse momentum smearing that arises in the final state hadronization process. It also describes the formation of a colorless hadronic state by emission of soft-gluon radiation. The SFs are related to the LDMEs of NRQCD, although the relation is not yet known in detail (the relation for large transverse momenta can be obtained from matching in the way described in \cite{Boer:2020bbd}). 
In this paper we will only consider the leading order relation, which means that we only consider Color Octet (CO) SFs and LDMEs. In the process $ep\to e'[Q\bar{Q}]X$ at leading order (LO) in $\alpha_{s}$ the heavy quark-antiquark will be in a CO state and CO LDMEs will dominate. Beyond LO also Color Singlet (CS) LDMEs will contribute, but as we will see, the uncertainty in the prediction from the uncertainty of the CO LDMEs is too large to necessitate inclusion of higher order corrections. As mentioned, another source of uncertainty is from the nonperturbative part of the TMD evolution expressions, for which we will estimate a reasonable range. Given these uncertainties, we can draw only qualitative conclusions on the dependence of the asymmetry on transverse momentum and how that differs from earlier results in the literature \cite{Mukherjee:2016qxa,Bacchetta:2018ivt,Kishore:2018ugo}. Nevertheless, the results are promising regarding future measurements of the asymmetry at the EIC, as it could become quite large. 

The paper is organized as follows. In Section \ref{Sec2}, we briefly review the TMD description of the process and introduce the SFs. In Section \ref{Sec3}, we explain the TMD evolution formalism that is employed for the numerical evaluations. Especially, we discuss the explicit form of the nonperturbative in more detail. In Section \ref{Sec4} we present our results on the azimuthal $\cos 2 \phi$ asymmetry, and finally conclude and discuss our findings in Section \ref{Sec5}.

\section{$J/\psi$ production within TMD factorization}
\label{Sec2}

\subsection{Leading order TMD description of the process}
We study the semi-inclusive process
\begin{align}
    e(l)+p(P_{h})\to e(l')+J/\psi(P)+X\hspace{1mm},
\end{align}
and we consider all particles to be unpolarized. The kinematic variables for this process are given by
\begin{align}
x_{B} = \frac{Q^{2}}{2 P_{h}\cdot q}\hspace{1mm}, \quad y = \frac{P_{h}\cdot q}{P_{h}\cdot l}\hspace{1mm}, \quad  z= \frac{P_{h}\cdot P}{P_{h}\cdot q}\hspace{1mm},
\end{align}
with $q=l-l'$ and $q^{2}=-Q^{2}$. The partonic subprocess contributing to this reaction at leading order (LO) in $\alpha_{s}$ is $\gamma^{*}(q)+g(p)\to Q\bar{Q}[n](P)$, with $n=^{2S+1}L_J^{(8)}$. The spectroscopic notation denotes that the produced $Q\bar{Q}$ pair will form a bound quarkonium state with spin $S$, orbital angular momentum $L$ and total angular momentum $J$. This transition from the $Q\bar{Q}$ pair to a bound quarkonium state is commonly described in terms of NRQCD LDMEs. At this LO only the CO configuration, indicated with the superscript $(8)$, will contribute. At order $\alpha_{s}^{2}$ also CS states contribute. Although the CS contribution is suppressed relatively to the CO by a perturbative coefficient of the order $\alpha_{s}/\pi$, according to the NRQCD scaling rules the relevant CO LDMEs $\langle 0|\OP_{8}^{J/\psi}(^{1}S_{0})|0\rangle$ and $\langle 0|\OP_{8}^{J/\psi}(^{3}P_{J})|0\rangle$ with $J=1,2,3$ are suppressed compared to $\langle 0|\OP_{1}^{J/\psi}(^{3}S_{1})|0\rangle$ by a factor of order $v^{3}$ and $v^{4}$, respectively \cite{Bodwin:2005hm}. Taken together, the CO configuration is enhanced with respect to the CS by a factor $v^{3}\pi/\alpha_{s}$ which for the smallest $Q$ value that we consider in this paper ($Q =3$ GeV) is around 2 and grows with increasing $Q$. Furthermore, it is known that at large $z$, the CS term becomes negligible \cite{Fleming:1997fq}, and $z$ is fixed to 1 in our analysis. For these reasons we solely consider CO contributions. 

\begin{figure}[hbt]
\centering
    \includegraphics[width=14cm]{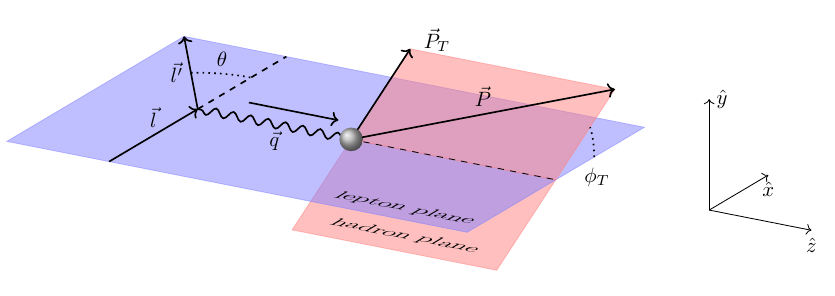}
\caption{Visualization of the azimuthal angle $\phi_{T}$, the scattering angle $\theta$, and the lepton and hadron scattering planes for the process $e(l)+p(P_{h})\to e(l')+J/\psi(P)+X$.}
\label{schematic}
\end{figure}

The reference frame for this process is chosen such that both the virtual photon from the electron and the incoming proton move along the $\hat{z}$-axis. The azimuthal angle $\phi_{T}$ of the quarkonium transverse momentum is defined with respect to the lepton scattering plane, i.e.\ $\phi_{l}=\phi_{l'}=0$. Moreover, the scattered electron has a scattering angle $\theta$ defined in this plane. In Figure \ref{schematic} a schematical setup of the reaction is shown. Because $z=1$, \begin{align}
    x = x_{B}\bigg(1+\frac{M^{2}}{Q^{2}}\bigg)\hspace{1mm},
\end{align}
where $M$ denotes the $J/\psi$ mass. In TMD factorization the differential cross section is schematically written as: 
\begin{align}
d \sigma \sim \int dx \hspace{1mm} d^{2}{\bf{p}}_{T} \hspace{2mm} \delta^{4}(p+q - P) \hspace{2mm} L(l,q) \hspace{1mm} \Gamma_{g}(x, {\bf{p}}_{T}) \hspace{1mm} \A(q,p) \hspace{1mm} \A^{*}(q,p) \hspace{1mm},
\end{align}
where ${\bf{p}}_{T}^2=-p_{T}^{2}$ denotes the transverse component of the gluon momentum, $L$ the leptonic tensor, $\Gamma_{g}$ the gluon correlator and $\A$ the partonic scattering amplitude. Subsequently, the correlator is parametrized at leading order in terms of two TMDs \cite{Mulders:2000sh}, with $f_{1}^{g}(x, {\bf{p}}_{T})$, the unpolarized gluon distribution and $h_{1}^{\perp g}(x, {\bf{p}}_{T})$, the linearly polarized gluon distribution. To ensure we can apply TMD factorization, only the kinematic region is considered in which the transverse momentum $P_{T}$ of the $J/\psi$ is small compared to the virtuality of the photon $Q$. Usually $P_{T} < Q/2$ is considered as the range of validity for the TMD region, but we will see that this is too optimistic in the present case. 

%\vspace{2mm}
%\vspace{-4mm}

Without taking into account smearing effects in the transition from the $Q\bar{Q}$ pair to a bound quarkonium state, the cross section is written in terms of TMDs and NRQCD LDMEs as (see e.g.\ \cite{Bacchetta:2018ivt}, where ${\bf q}_{T} \equiv {\bf P}_{T}$)
\begin{align}
\frac{d\sigma (J/\psi)}{dx_{B} \hspace{1mm} dy \hspace{1mm} d^{2}{\bf{q}}_{T}} =
\N \bigg[ {\cal A} \hspace{1mm} f_{1}^{g}(x, {\bf{q}}_{T}^{2}) + \frac{{\bf{q}}_{T}^{2}}{M_{h}^{2}}\hspace{1mm} {\cal B} \hspace{1mm}h_{1}^{\perp g} (x, {\bf{q}}_{T}^{2})\hspace{1mm} \text{cos} \hspace{1mm}(2\phi_{T}) \bigg]  \hspace{1mm},
\label{sigma}
\end{align} 
with the normalization factor 
\begin{align*}
\N = (2 \pi)^{2} \frac{\alpha^{2} \alpha_{s} e_{c}^{2}}{y Q^{2} M (M^{2}+Q^{2})} \hspace{1mm},
\end{align*}
where $e_{Q}$ is the fractional electric charge of the quark and $M_{h}$ the mass of the proton. The explicit forms of the prefactors are
\begin{align}
& {\cal A}=[(1-y)^{2}+1] \A_{U+L}^{\gamma^{*}g\to J/\psi} - y^{2}\A_{L}^{\gamma^{*}g\to J/\psi} \hspace{1mm}, \label{AF} \\
& {\cal B}= (1-y)\B_{T}^{\gamma^{*}g\to J/\psi} \hspace{1mm} \label{BF},
\end{align}
where $y$ is the inelasticity and the superscripts $U+L$, $L$ and $T$ refer to the specific polarization of the photon \cite{Pisano:2013cya, Brodkorb:1994de}. Employing heavy-quark spin symmetry relations \cite{Bodwin:1994jh},  one finds in terms of CO LDMEs \cite{Bacchetta:2018ivt}:
\begin{align}
 \A_{U+L}^{\gamma^{*}g\to J/\psi} &= \langle 0| \OP_{8}^{J/\psi}(^{1}S_{0})|0 \rangle + \frac{12}{N_{c}} \frac{7M^{2}+3Q^{2}}{M^{2}(M^{2}+Q^{2})} \langle 0| \OP_{8}^{J/\psi}(^{3}P_{0})|0 \rangle \hspace{1mm},\\
 \A_{L}^{\gamma^{*}g\to J/\psi} &= \frac{96}{N_{c}} \frac{Q^{2}}{(M^{2}+Q^{2})^{2}} \hspace{1mm} \langle 0| \OP_{8}^{J/\psi}(^{3}P_{0})|0 \rangle \hspace{1mm},\\
 \B_{T}^{\gamma^{*}g\to J/\psi} &= -\langle 0| \OP_{8}^{J/\psi}(^{1}S_{0})|0 \rangle + \frac{12}{N_{c}} \frac{3M^{2}-Q^{2}}{M^{2}(M^{2}+Q^{2})} \langle 0| \OP_{8}^{J/\psi}(^{3}P_{0})|0 \rangle \hspace{1mm}.
 \label{prefFs}
\end{align}
Taking into account evolution in the above result for the cross section requires consideration of the more general TMD factorization expression including SFs.

\subsection{Smearing effects and shape functions}
As mentioned, the TMD factorized expressions have to take into account final state smearing effects that are encoded in the SF $\Delta^{[n]}$ \cite{Echevarria:2019ynx, Fleming:2019pzj}. This nonperturbative hadronic quantity describes the transition from the $Q\bar{Q}$ pair to a bound quarkonium state, not only the formation of the bound state, but also the soft-gluon radiation required to produce a final state hadron in the CS state, which generally will change the momentum of the quarkonium. Including the SFs in the cross section becomes
\begin{align}
\frac{d\sigma (J/\psi)}{dx_{B} \hspace{1mm} dy \hspace{1mm} d^{2}{\bf{q}}_{T}}=
\N \bigg[\sum_n {\cal A}^{[n]} \hspace{1mm} \Co[f_{1}^{g}\hspace{1mm}\Delta^{[n]}]+2 \hspace{1mm} \sum_n {\cal B}^{[n]} \hspace{1mm}\Co[wh_{1}^{\perp g}\hspace{1mm}\Delta^{[n]}_{h}]\hspace{1mm}\text{cos}\hspace{1mm} (2\phi_{T})\bigg] \hspace{1mm},
\end{align}
where we have introduced the transverse momentum convolutions:
\begin{align}
&\Co[f_{1}^{g}\hspace{1mm}\Delta^{[n]}](x,{\bf{q}}_{T}^{2}) = \int d^{2}{\bf{p}}_{T} \int d^{2}{\bf{k}}_{T} \hspace{1mm} \delta^{2}({\bf{p}}_{T}+{\bf{k}}_{T}-{\bf{q}}_{T}) \hspace{1mm} f_{1}^{g}(x,{\bf{p}}_{T}^{2}) \hspace{1mm} \Delta^{[n]}({\bf{k}}_{T}^{2}) \hspace{1mm},\\
&\Co[wh_{1}^{\perp g}\hspace{1mm}\Delta^{[n]}_{h}](x,{\bf{q}}_{T}^{2}) = \int d^{2}{\bf{p}}_{T} \int d^{2}{\bf{k}}_{T} \hspace{1mm} \delta^{2}({\bf{p}}_{T}+{\bf{k}}_{T}-{\bf{q}}_{T}) \hspace{1mm} w({\bf{p}}_{T},{\bf{k}}_{T}) \hspace{1mm} h_{1}^{\perp g}(x,{\bf{p}}_{T}^{2}) \hspace{1mm} \Delta^{[n]}_{h}({\bf{k}}_{T}^{2})\hspace{1mm},
\end{align}
with the transverse momentum dependent weight function 
\begin{align}
w({\bf{p}}_{T},{\bf{k}}_{T}) = \frac{1}{2M_{h}^{2}{\bf{q}}_{T}^{2}}[2({\bf{p}}_{T} \cdot {\bf{q}}_{T})^{2} - {\bf{p}}_{T}^{2} {\bf{q}}_{T}^2] \hspace{1mm}.
\end{align}
The SF can be thought of as a generalization of the LDMEs in collinear factorization. It is expected that the shape functions are proportional to the LDMEs, also beyond LO: 
\begin{align}
\Delta^{[n]}_{(h)} ({\bf{k}}_{T}^{2}) \equiv \langle 0| \OP(n)|0 \rangle \, \Delta_{(h)} ({\bf{k}}_{T}^{2})\hspace{1mm},
\label{univeralSFs}
\end{align}
for some universal SFs $\Delta ({\bf{k}}_{T}^{2})$ and $\Delta_{h} ({\bf{k}}_{T}^{2})$ that could in principle be unequal. In absence of smearing $\Delta ({\bf{k}}_{T}^{2})= \Delta_{h} ({\bf{k}}_{T}^{2})= \delta^{2}({\bf{k}}_{T})$, the convolutions reduce to products of an LDME with a gluon TMD and Eqs.\ (\ref{sigma})-(\ref{prefFs}) are recovered. This is the simplification we will adopt, but only after including effects from TMD evolution.   

The process under investigation has an azimuthal asymmetry
\begin{align}
\langle \text{cos}\hspace{1mm} 2\phi_{T} \rangle = \frac{\int d\phi_{T} \hspace{1mm} \text{cos}\hspace{1mm} 2\phi_{T} \hspace{1mm} d\sigma}{\int d\phi_{T} \hspace{1mm} d\sigma} = \frac{\sum_n {\cal B}^{[n]} \hspace{1mm}\Co[wh_{1}^{\perp g}\hspace{1mm}\Delta^{[n]}_{h}]}{\sum_n {\cal A}^{[n]} \hspace{1mm} \Co[f_{1}^{g}\hspace{1mm}\Delta^{[n]}]}
\hspace{1mm}.
\end{align}
Adopting Eq.\ (\ref{univeralSFs}), this becomes 
\begin{align}
\langle \text{cos}\hspace{1mm} 2\phi_{T} \rangle = \frac{\sum_n {\cal B}^{[n]}\, \langle 0| \OP(n)|0 \rangle }{\sum_n {\cal A}^{[n]}\, \langle 0| \OP(n)|0 \rangle } \cdot R = \frac{{\cal B}}{{\cal A}} \cdot R
\hspace{1mm},
\end{align}
where ${\cal A}$ and ${\cal B}$ are given in Eqs.\ (\ref{AF}) and (\ref{BF}), 
and 
\begin{align}
    R = \frac{\Co[wh_{1}^{\perp g}\hspace{1mm}\Delta_{h}]}{\Co[f_{1}^{g}\hspace{1mm}\Delta]} \hspace{1mm}.
\end{align}
Before continuing, we make a few remarks on the prefactor ${\cal B}/{\cal A}$. It turns out that ${\cal B}/{\cal A}$ depends very strongly on the specific set of LDMEs adopted. Also, since $|\langle \text{cos}\hspace{1mm} 2\phi_{T} \rangle | \leq 1$ and $|R| \leq 1$, the LDMEs must be such that $|{\cal B}/{\cal A}| \leq 1$ and ${\cal A} \geq 0$, which are important constraints to impose on LDME extractions at EIC. These constraints are not satisfied by the BK set for $Q^2 \simorderl 2.5 M^2$, but that seems a problem of applying LDMEs that were obtained from an NLO analysis in a LO analysis. Finally, we note that ${\cal B}/{\cal A}$ vanishes in the limit $y \to 1$ when the virtual photon is longitudinally polarized and maximizes when $y\to 0$.

\section{TMD evolution implementation}
\label{Sec3}
Beyond tree level, the TMDs and SFs, become scale dependent which governs the TMD evolution \cite{Collins:1981uk}. Implementing TMD evolution is more easily done in impact parameter space, where convolutions become simple products. In general, we can write
\begin{align}
\frac{d\sigma (J/\psi)}{dx_{B} \hspace{1mm} dy \hspace{1mm} d^{2}{\bf{q}}_{T}} = \int d^{2}{\bf{b}}_{T} \hspace{1mm} e^{-i{\bf{b}}_{T} \cdot {\bf{q}}_{T}} \hspace{1mm} \hat{W}({\bf{b}}_{T}, Q) + \OP({\bf{q}}_{T}^{2}/Q^{2}) \hspace{1mm},
\end{align}
where $b_{T}$ is the conjugate of momentum $q_{T}$. $\hat{W}$ consist of three factors:
\begin{align}
\hat{W}({\bf{b}}_{T}, Q) =  \hspace{1mm} \hat{A}(x,{\bf{b}}_{T}; \zeta_{A}, \mu) \hspace{1mm} \hat{B}({\bf{b}}_{T}; \zeta_{B}, \mu)\hspace{1mm}\Ha(Q; \mu) \hspace{1mm}.
\end{align}
Here $\Ha$ denotes the hard part, $\hat{A}$ and $\hat{B}$ the Fourier transformed TMD and SF, respectively, $\mu$ the renormalization scale and $\zeta_{A/B}$ the rapidity variables. 
The latter arises due to the required regularization of lightcone divergences from lightlike Wilson lines \cite{Collins:1981uk}, although in this case there are no such divergences associated to the SF \cite{Sun:2012vc,Echevarria:2019ynx}. The natural choice to minimize large logarithms in $\zeta_{A/B}$ will be $\zeta_A = Q^2, \zeta_B=1$, similar as in \cite{delCastillo:2021znl} for open heavy quark pair production in electron-proton collisions. 

The Fourier transforms of $f_{1}^{g}$, $h_{1}^{\perp g}$ and $\Delta^{[n]}_{(h)}$ are defined as follows:
\begin{align}
\hat{f}_{1}^{g}(x,{\bf{b}}_{T}^{2}) &\equiv \int d^{2}{\bf{p}}_{T} \hspace{1mm} e^{i {\bf{b}}_{T} \cdot {\bf{p}}_{T}} \hspace{1mm}f_{1}^{g}(x,{\bf{p}}_{T}^{2})\hspace{1mm}, \\
\hat{h}_{1}^{\perp g}(x,{\bf{b}}_{T}^{2}) &\equiv \int d^{2}{\bf{p}}_{T} \hspace{1mm} \frac{({\bf{b}}_{T}\cdot{\bf{p}}_{T})^{2} - \frac{1}{2} {\bf{b}}_{T}^{2}{\bf{p}}_{T}^{2}}{{\bf{b}}_{T}^{2}M_{h}^{2}}\hspace{1mm}e^{i {\bf{b}}_{T} \cdot {\bf{p}}_{T}} \hspace{1mm}h_{1}^{\perp g}(x,{\bf{p}}_{T}^{2})\hspace{1mm}, \\
\hat{\Delta}^{[n]}_{(h)}({\bf{b}}_{T}^{2}) &\equiv \int d^{2}{\bf{k}}_{T} \hspace{1mm} e^{i {\bf{b}}_{T} \cdot {\bf{k}}_{T}} \hspace{1mm}\Delta^{[n]}_{(h)}({\bf{k}}_{T}^{2})\hspace{1mm}.
\end{align}
In terms of these functions the convolutions can be written as:
\begin{align}
\Co[f_{1}^{g}\hspace{1mm}\Delta^{[n]}] &= \int_{0}^{\infty} \frac{db_{T}}{2\pi} \hspace{1mm} b_{T}\hspace{1mm} J_{0}(b_{T}q_{T})\hspace{1mm}\hat{f}_{1}^{g}(x,{\bf{b}}_{T}^{2}) \hspace{1mm} \hat{\Delta}^{[n]}({\bf{b}}_{T}^{2})\hspace{1mm}, \\
\Co[wh_{1}^{\perp g}\hspace{1mm}\Delta^{[n]}_{h}] &= - \hspace{1mm} \int_{0}^{\infty} \frac{db_{T}}{2 \pi}  \hspace{1mm} b_{T} \hspace{1mm} J_{2}(b_{T}q_{T}) \hspace{1mm} \hat{h}_{1}^{\perp g}(x,{\bf{b}}_{T}^{2}) \hspace{1mm} \hat{\Delta}^{[n]}_{h}({\bf{b}}_{T}^{2}) \hspace{1mm},
\end{align}
where we suppressed the dependence on $\zeta$ and $\mu$. 

The TMDs obey the Collins-Soper and Renormalization Group equations with respect to the scale parameters $\zeta$ and $\mu$ \cite{Collins:1981uk}. These equations can be used to evolve the TMDs from high to low scales \cite{Collins:1981uk,Aybat:2011zv}:
\begin{align}
& \hat{A}(x,{b}_{T};\zeta, \mu) = e^{-S_{A}(b_{T};\zeta,\zeta_{0},\mu,\mu_{0})} \hspace{1mm} \hat{A}(x,{b}_{T};\zeta_{0}, \mu_{0}) \hspace{1mm},
\end{align}
with Sudakov factor $S_{A}$:
\begin{align}
S_{A}(b_{T};\zeta,\zeta_{0},\mu,\mu_{0}) = -\frac{1}{2}\hat{K}(b_{T}; \mu_{0}) \hspace{1mm} \text{ln} \hspace{1mm} \frac{\zeta}{\zeta_{0}} - \int_{\mu_{0}}^{\mu} \frac{d\mu'}{\mu'} \hspace{1mm} \bigg[\gamma(\alpha_s(\mu');1) -\frac{1}{2}\gamma_{K}(\alpha_s(\mu'))\hspace{1mm} \text{ln} \hspace{1mm} \frac{\zeta}{\mu'^{2}}\bigg] \hspace{1mm}.
\end{align}
While the renormalization scale $\mu$ in the hard scattering part $\Ha$ should be set to $\mu \sim Q$ to avoid large logarithms of $\mu/Q$, the TMDs should be evaluated at much lower scale in order to avoid large logarithms of $\mu/M_h$ or $\mu/\Lambda_{\text{QCD}}$. Instead of selecting a fixed, low (but still perturbative) scale for the TMDs, it 
is common to take $\sqrt{\zeta_{0}} \sim \mu_{0} \sim \mu_{b} \equiv  b_{0}/b_{T}$ and make sure in the calculation that
$\mu_b \leq Q$. The Sudakov factor then expresses the resummation of logarithms in $\mu_b/Q$.  

In leading order in $\alpha_s$ the perturbative expression for the above Sudakov factor can be written as \cite{Collins:1981uk,Aybat:2011zv} 
\begin{align}
S_{A}(b_{T};Q,\mu_{b}) &= \frac{1}{2}\frac{C_{A}}{\pi} \int_{\mu_{b}^{2}}^{Q^{2}} \frac{d\mu'^{2}}{\mu'^{2}} \hspace{1mm} \alpha_s(\mu') \hspace{1mm} \bigg[\text{ln} \hspace{1mm} \frac{Q^{2}}{\mu'^{2}} -\bigg(\frac{11-2n_{f}/C_{A}}{6}\bigg)\bigg] + \OP(\alpha_{s}^{2})\hspace{1mm}. 
\label{SAover2}
\end{align}
Adopting the scale $\mu \sim Q$ in the SF is also expected to lead to large logarithms in $Q/\mu_b$, which should also be resummed. This can again be done using a Renormalization Group equation, leading to a contribution to the overall Sudakov factor at the single log level. It can be incorporated by changing $S_A$ into:
\begin{align}
S_{A}(b_{T};Q,\mu_{b}) &= \frac{1}{2}\frac{C_{A}}{\pi} \int_{\mu_{b}^{2}}^{Q^{2}} \frac{d\mu'^{2}}{\mu'^{2}} \hspace{1mm} \alpha_s(\mu') \hspace{1mm} \bigg[\text{ln} \hspace{1mm} \frac{Q^{2}}{\mu'^{2}} -\bigg(\frac{11-2n_{f}/C_{A}}{6}+\text{B}_{\text{CO}}\bigg)\bigg] + \OP(\alpha_{s}^{2})\hspace{1mm}.
\label{SA+SG}
\end{align}
The constant $\text{B}_{\text{CO}}$ from the soft-gluon radiation is not derived for the current process yet. However, since $\Delta_{(h)} ({\bf{k}}_{T}^{2})$ in Eq.\ (\ref{univeralSFs}) is expected to be universal, we take for the present case of CO $J/\psi$ production $\text{B}_{\text{CO}}=1$ as obtained in $pp\to \eta_c\,X$ for CO $\eta_c$ production \cite{Sun:2012vc}. This is also consistent with recent results presented in \cite{Echevarria:2022} obtained within the SCET formalism.

Including the one-loop running of $\alpha_{s}$, one can preform the $\mu$ integral explicitly:
\begin{align}
S_{A}(b_{T}; Q, \mu_{b}) = & -\frac{1}{2}\frac{36}{33-2 n_{f}}\bigg[\text{ln} \hspace{1mm} \frac{Q^{2}}{\mu_{b}^{2}} +  \text{ln}\hspace{1mm} \frac{Q^{2}}{\Lambda_{\text{QCD}}^{2}}  \hspace{1mm} \text{ln} \hspace{1mm} \bigg(1- \frac{\text{ln} \hspace{1mm} (Q^{2}/\mu_{b}^{2})}{\text{ln} \hspace{1mm} (Q^{2}/\Lambda_{\text{QCD}}^{2})} \bigg) \nonumber \\
& + \bigg(\frac{11-2n_{f}/C_{A}}{6} +\text{B}_{\text{CO}}\bigg)\hspace{1mm} \text{ln} \hspace{1mm} \bigg(\frac{\text{ln} \hspace{1mm} (Q^{2}/\Lambda_{\text{QCD}}^{2})}{\text{ln} \hspace{1mm} (\mu_{b}^{2}/\Lambda_{\text{QCD}}^{2})} \bigg)\bigg] + \OP(\alpha_{s}^{2})\hspace{1mm}.
\end{align}
Note that $S_{A}$ is spin independent, and thus the same for all convolutions. 

We note that the perturbative Sudakov factor for our process is quite different from the one of \cite{Boer:2020bbd}, where it is assumed that the SF also leads to a factor $S_A$ given in Eq.\ (\ref{SAover2}), like for TMD fragmentation functions. This leads to a larger Sudakov factor and as a consequence to more Sudakov suppression. We then find smaller values for the asymmetry, but the shape of the asymmetry is quite similar.

The above perturbative expression for the Sudakov factor is valid in the region $b_0/Q \leq b_T \leq b_{T,\text{max}}$. The lower limit is the point beyond which $\mu_b$ becomes larger than $Q$, such that the Sudakov integral flips sign. 
The upper limit marks the point where perturbation theory starts to fail, which is not exactly known. It is common to take $b_{T,\text{max}} = 0.5\hspace{1mm}\text{GeV}^{-1}$ or $b_{T,\text{max}} = 1.5\hspace{1mm}\text{GeV}^{-1}$ in phenomenological analyses. 

The two limits to the $b_T$ integration are implemented in different ways. One way to ensure $b_0/Q \leq b_T$ is to 
consider the replacement \cite{Boer:2014tka}
\begin{align}
\mu_{b}\to \mu_{b}'=\frac{Qb_{0}}{Qb_T+b_{0}}\hspace{1mm}.
\label{mubprime}
\end{align}
in the Sudakov factor, which effectively boils down to a different resummation: in logarithms of $\mu_b'/Q$ rather than $\mu_b/Q$. For consistency this is then also the scale one should use in the TMDs and SFs. 
In this way the perturbative expression for $S_A$ is valid for all $b_T \leq b_{T,\text{max}}$. Often a slightly different replacement is considered \cite{Collins:2016hqq}
\begin{align}
\mu_{b}\to \tilde{\mu}_{b}'=\frac{b_{0}}{\sqrt{b^2+b_{0}^2/Q^2}} \hspace{1mm}.
\label{mubprimeC}
\end{align}
However, we will use the $\mu_{b}'$ replacement for our predictions. We will comment on this choice below.

Different ways to ensure that $b_T \leq b_{T,\text{max}}$ in the perturbative expression have been employed, but the most common one is the $b_{T}^{*}$-method \cite{Collins:1984kg}: 
\begin{align}
 b_{T}^{*}(b_{T}) = \frac{b_{T}}{\sqrt{1+\left(b_{T}/b_{T,\text{max}}\right)^{2}}} \hspace{1mm}.
\label{bstar}
\end{align}
This is introduced in the following way:
\begin{align}
\hat{W}(b_{T})\equiv \hat{W}(b_{T}^{*})e^{-S_{NP}}\hspace{1mm},
\end{align}
where for $\hat{W}(b_{T}^{*})$ one can always use the perturbative expression for the Sudakov factor and the nonperturbative Sudakov factor $S_{NP}$ makes up for the difference to the real $\hat{W}(b_{T})$. 
This factor can only be extracted from data and is in principle different for different convolutions. 
In addition, $S_{NP}$ depends on the prescriptions used to separate the perturbative and nonperturbative components inside the convolution. There are different parameterizations used in the literature, but typically it is chosen to be a Gaussian. In general it is $Q$ dependent and of the form \cite{Collins:1984kg}:
\begin{align}
S_{NP}(b_{T};Q) = \text{ln} \hspace{1mm} \bigg( \frac{Q}{Q_{NP}}\bigg) g(b_{T})+ g_{\text{TMD}}(b_{T})+ g_{SF}(b_{T})\hspace{1mm},
\end{align}
where $Q_{NP}$ is a parameter with the dimension of mass, that typically is chosen to be (near) the smallest scale at which perturbation theory is expected to be valid; of course, any change in $Q_{NP}$ can be compensated for by a change in the other terms. The specific form of $S_{NP}$ will be discussed in the next section. General constraints are that $\text{exp} \hspace{1mm}(-S_{NP})$ should be unity at $b_{T}=0$, and it should smoothly vanish at large $b_{T}$, in order to exclude contributions from (far) outside the proton. It also guarantees convergence of the convolutions. 

Implementing both (\ref{mubprime}) and (\ref{bstar}) should be done in the right order, i.e.  
\begin{align}
\mu_{b}\to \mu_{b}'=\frac{Qb_{0}}{Qb_T+b_{0}} \to \mu_{b^*}' = \frac{Qb_{0}}{Qb_T^*+b_{0}}\hspace{1mm},
\label{primestar}
\end{align}
or similarly for $\tilde{\mu}_{b}'$.
In this way one ensures that $S_A(0)=0$. This is also the scale that one should adopt in the TMDs and SFs. 

If we take all the above into account, the convolutions read
\begin{align}
\Co[f_{1}^{g}\hspace{1mm}\Delta^{[n]}]&= \int_{0}^{\infty} \frac{db_{T}}{2\pi} \hspace{1mm} b_{T}\hspace{1mm} J_{0}(b_{T}q_{T})\hspace{1mm}e^{-S_{A}(b_{T}^{*};Q,\mu_{b^*}')} e^{-S_{NP}(b_{T};Q)} \hspace{1mm} \hat{f}_{1}^{g}(x,b_{T}^{*}) \hspace{1mm} \hat{\Delta}^{[n]}(b_{T}^{*}) \hspace{1mm},\\
\Co[wh_{1}^{\perp g}\hspace{1mm}\Delta^{[n]}_{h}] &= - \hspace{1mm} \int_{0}^{\infty} \frac{db_{T}}{2\pi} \hspace{1mm} b_{T} \hspace{1mm} J_{2}(b_{T}q_{T})\hspace{1mm} e^{-S_{A}(b_{T}^{*};Q,\mu_{b^*}')} e^{-S_{NP}(b_{T};Q)} \hspace{1mm} \hat{h}_{1}^{\perp g}(x,b_{T}^{*}) \hspace{1mm} \hat{\Delta}^{[n]}_{h}(b_{T}^{*}) \hspace{1mm},
\end{align}
where we suppressed the dependence on the scale $\sqrt{\zeta} = \mu = \mu_{b^*}'$ in the TMDs and SFs.  
At this perturbative scale, the $b_T$ dependence of the TMDs and SFs can be calculated. For the TMDs these ``perturbative tails'' are \cite{Sun:2011iw}:
\begin{align}
\hat{f}_{1}^{g}(x,b_{T};\mu_{b}^{2},\mu_{b})&=f_{g/P}(x;\mu_{b})+ \OP(\alpha_{s}) + \OP(b_{T}\Lambda_{\text{QCD}}) \hspace{1mm},\\
\hat{h}_{1}^{\perp g}(x,b_{T};\mu_{b}^{2},\mu_{b}) &=-\frac{\alpha_{s}(\mu_{b})}{\pi}\int_{x}^{1} \frac{dx'}{x'} \hspace{1mm} \bigg(\frac{x'}{x}-1\bigg)\hspace{1mm} \bigg\{ C_{A} \hspace{1mm}f_{g/P}(x';\mu_{b})+ C_{F} \sum_{i=q,\bar{q}}  f_{i/P}(x';\mu_{b}) \bigg\} \nonumber \\ & + \OP(\alpha_{s}^{2}) + \OP(b_{T}\Lambda_{\text{QCD}}) \hspace{1mm}.\label{h1perptail}
\end{align}
One sees that both expressions are determined by the collinear distributions $f_{i/P}$, but start at different orders in $\alpha_{s}$, the reason being that there is no collinear version of $h_{1}^{\perp g}$.

The leading order perturbative transverse momentum tail of $\hat{\Delta}^{[n]}$ has been studied in \cite{Boer:2020bbd}, where it is observed that the tail of $\Delta^{[n]}_{h}$ would require a study at higher order in $\alpha_s$. However, the Fourier transformed functions, i.e.\ the perturbative small-$b$ expressions of the SFs, both start at order $\alpha_s^0$, in contrast to the TMDs discussed above. This order $\alpha_s^0$ contribution stems from the small transverse momentum part, irrespective of whether it is the simple $\Delta ({\bf{k}}_{T}^{2})= \Delta_{h} ({\bf{k}}_{T}^{2})= \delta^{2}({\bf{k}}_{T})$ expression or more realistic smeared-out versions of $\Delta ({\bf{k}}_{T}^{2})$ and $\Delta_{h} ({\bf{k}}_{T}^{2})$. Given the uncertainty in the $\alpha_s$ corrections to the tails of the SFs, here we will simply adopt the LO description in terms of the LDMEs.

In this section we have discussed all perturbative ingredients to perform TMD evolution numerically at leading order in $\alpha_{s}$. In order to make numerical predictions we need to discuss the nonperturbative Sudakov factor in greater detail. As this factor is unknown for gluons, we need to estimate the uncertainty this introduces in the predictions.

\subsection{Nonperturbative Sudakov factor and error estimation}

A parameterization for $S_{NP}$ was obtained from fits to low energy SIDIS data as well as higher energy DY and $Z$ boson production data \cite{Aybat:2011zv}
\begin{align}
S_{NP}(b_{T};Q) = \bigg[g_{1} \hspace{1mm} \text{ln} \frac{Q}{2Q_{NP}}+g_{2}\bigg(1+2g_{3}\hspace{1mm} \text{ln}\frac{10xx_{0}}{x_{0}+x}\bigg)\bigg]b_{T}^{2}\hspace{1mm},
\label{SNPAR}
\end{align}
with $g_{1} = 0.184\hspace{1mm}\text{GeV}^{2}$, $g_{2} = 0.201\hspace{1mm}\text{GeV}^{2}$, $g_{3} =- 0.129\hspace{1mm}\text{GeV}^{2}$, $x_{0} = 0.009$, $Q_{NP} = 1.6\hspace{1mm}\text{GeV}$ and $b_{T,\text{max}} = 1.5\hspace{1mm}\text{GeV}^{-1}$. We take this expression as our starting point for two fixed small $x$ values and 
Casimir scaled (i.e.\ multipied by a factor $C_{A}/C_{F}$) in order to apply to gluons:
\begin{align}
S_{NP}(b_{T};Q) = \bigg[ A \hspace{1mm} \text{ln} \frac{Q}{Q_{NP}}+ B(x)\bigg]b_{T}^{2}\hspace{1mm}, \hspace{2mm} \text{with} \hspace{2mm} A=\frac{C_A}{C_F}g_1 = 0.414\hspace{1mm} \text{GeV}^{2} \hspace{1mm}.
\label{SNPABform}
\end{align}
The $B$-term can be thought of as related to the intrinsic transverse momentum of the TMDs, which generally is $x$ dependent. By Fourier transforming a simple Gaussian dependence for $f_{1}^{g}(x,{\bf{p}}_{T}^{2})\propto \text{exp}[-{\bf{p}}_{T}^{2}/\langle p_{T}^{2}\rangle]$, we obtain $B \approx \langle p_{T}^{2}\rangle/4$ (at $Q=Q_{NP}$). Of course, the $B$-term is not necessarily the same for the $h_{1}^{\perp g}(x,{\bf{p}}_{T}^{2})$ convolution, but for simplicity we do not distinguish between these cases. Matching Eqs.\ (\ref{SNPAR}) and (\ref{SNPABform}) for the $x$ values that we will consider yields the specific $B$ values shown in Table \ref{table1}.

\begin{table}[htb]
\centering
\caption{Values of the parameters $A$ and $B$ used in $S_{NP}$. Left the $A$ values are shown along with the corresponding $b_{T,\text{lim}}$ and $r$ determined at $Q = 12$ GeV. Right the $B$ values are shown determined by $x$.}
\label{table1}
\vspace{4mm}
\scalebox{1}{
\begin{tabular}{ c c | c }
\hline
 \rule{0pt}{12pt} $b_{T,\text{lim}}$ $(\text{GeV}^{-1})$ \ & $r \hspace{1mm} (\text{fm} \sim 1/(0.2 \hspace{1mm} \text{GeV}))$ & $A \hspace{1mm}(\text{GeV}^{2})$\\
\hline
\hline
2 \ & 0.2 & 0.80\\
4 \ & 0.4 & 0.20\\
8 \ & 0.8 & 0.05\\
\hline
\end{tabular}
\quad
\begin{tabular}{ c | c }
\hline
 \rule{0pt}{12pt} $x$ & $B \hspace{1mm}(\text{GeV}^{2})$ \\
\hline
\hline
$10^{-1}$ & 0.456\\
$10^{-2}$ & 0.521\\
$10^{-3}$ & 0.715\\
\hline
\end{tabular}}
\end{table}

To perform an error estimate and to assess the importance of $S_{NP}$ for the size of the convolutions and asymmetry, we will vary $A$ within the extreme limits following a previous study \cite{Scarpa:2019fol}, instead of taking the one quoted in Eq.\ (\ref{SNPABform}). The idea is that one expects the $\exp(-S_{NP})$ term to be non-negligible (here defined as being larger than $10^{-3}$) anywhere between $b_{T,\text{max}}$ and the charge radius of the proton. If the $\exp(-S_{NP})$ term becomes negligible around $b_{T,\text{max}}$ already, then there will be hardly any nonperturbative contribution outside the perturbative regime, which is not realistic. On the other hand, one does not expect significant contributions beyond the charge radius of the proton, offering a generous upper bound. To implement this range, we define a value $b_{T,\text{lim}}$ such that at large $Q$, where the $A$-term is dominant and the $B$-term can be neglected, $\text{exp}[- S_{NP}]$ becomes negligible, i.e.\ $\sim$ $10^{-3}$. Considering $b_{T,\text{lim}}$ as the diameter, since it is conjugate to $q_{T}$, $r$ is defined as the characteristic radius $r = b_{T,\text{lim}}/2$, that can be thought of as the range over which the interactions occur from the center of the proton. For three values of $b_{T,\text{lim}}$ or $r$ we determined the $A$ values at $Q = 12$ GeV, which are shown in Table \ref{table1}. Note that $A = 0.414\hspace{1mm} \text{GeV}^{2}$ from Eq.\ (\ref{SNPABform}) lies roughly in the middle of this range.

\begin{figure}[htb]
    \centering
    \includegraphics[width=15.5cm]{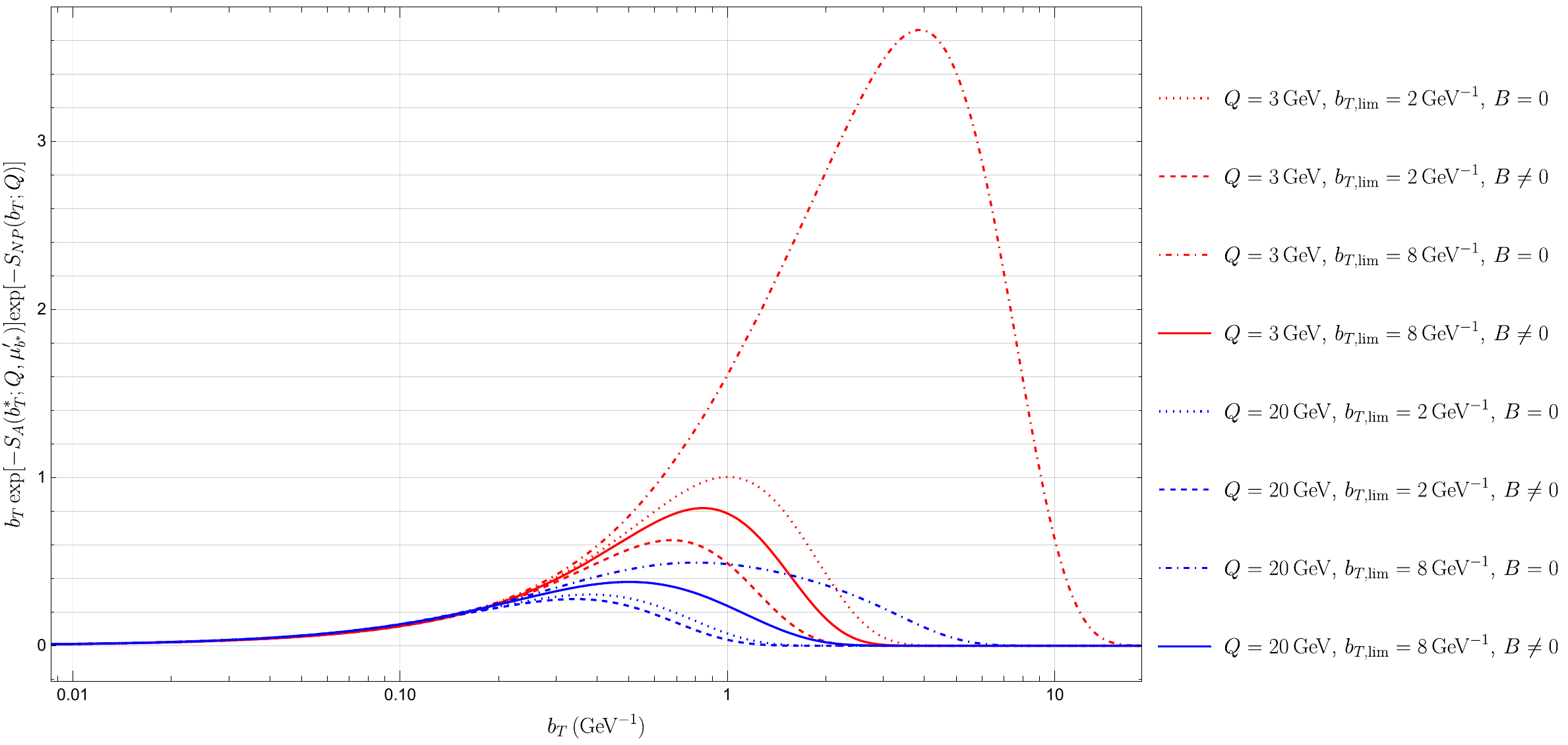}
    \caption{$b_{T}$ times the product of the Sudakov factors as a function of $b_{T}$ for $Q=3$ GeV and $Q=20$ GeV, which shows the importance of the extra $B$-term in $S_{NP}$ at small $Q$ and large $b_{T,\text{lim}}$, suppressing the unwanted contributions at large $b_{T}$. Here $b_{T,\text{max}} =1.5\hspace{1mm}\text{GeV}^{-1}$ and $B$ at $x=10^{-3}$ is used. 
    }
    \label{Snpeffect}
\end{figure}

We will include the $B$-term in $S_{NP}$ as it is needed for smaller $Q$ values and large $b_{T,\text{lim}}$ values. This is illustrated in Figure \ref{Snpeffect}: the product $\text{exp}[-S_{A}]\hspace{1mm}\text{exp}[-S_{NP}]$ at $Q=3$ GeV receives large perturbative contributions from $S_A$ at large $b_{T}$, all the way up to or even beyond the proton radius. This results in an upward bump at small $q_{T}$ in the convolutions, in particular in $\Co[wh_{1}^{\perp g}\hspace{1mm}\Delta_h^{[n]}]$. Moreover, at small $Q$ the limit of $B \to 0$ can result in a violation of the positivity bound for the convolutions: $\Co[wh_{1}^{\perp g}\hspace{1mm}\Delta^{[n]}_{h}] \leq \Co[f_{1}^{g}\hspace{1mm}\Delta^{[n]}]$. For larger $Q$ values, the curves with and without $B$-term lie increasingly closer to each other, as expected.

\section{Numerical predictions for the azimuthal asymmetry}
\label{Sec4}
For the numerical calculations we can use different sets of extractions of CO LDMEs for the $J/\psi$ case. These values are obtained from fits to Tevatron, RHIC and LHC data and summarized in Table \ref{table2}. Most of these results are obtained from next-to-leading order analyses, except for the SV set, which is based on a leading order calculation. The mass of the $J/\psi$ ($M$) and the charm quark ($m_c$) are taken to be $3.1$ GeV and $1.4$ GeV, respectively. 

\begin{table}[htb]
\centering
\caption{Numerical values of the LDMEs for $J/\psi$ production in units of $10^{-2} \hspace{1mm} \text{GeV}^{3}$.}
\label{table2}
\vspace{4mm}
\begin{tabular}{ l | c | c |}
& $\langle 0| \OP_{8}^{J/\psi}(^{1}S_{0})|0 \rangle$ & $\langle 0| \OP_{8}^{J/\psi}(^{3}P_{0})|0 \rangle / m_{c}^{2}$  \\
\hline
\hline
CMSWZ {\cite{Chao:2012iv}} &  $8.9 \pm 0.98$ & $0.56 \pm 0.21$  \\
\hline
SV {\cite{Sharma:2012dy}} &  $1.8 \pm 0.87$ & $1.8 \pm 0.87$ \\
\hline
BK {\cite{Butenschoen:2010rq}} &  $4.5 \pm 0.72$ & $-0.54 \pm 0.16$ \\
\hline
BCKL {\cite{Bodwin:2014gia}} &  $9.9 \pm 2.2$ & $0.49 \pm 0.44$ \\
\hline
\end{tabular}
\end{table}

The perturbative tails of the TMDs in terms of the collinear quark and gluon PDFs, are computed by using the MSTW2008LO PDF set \cite{Martin:2009iq}. Furthermore, we employ one-loop running of $\alpha_{s}$. To determine the QCD scale we matched $\alpha_{s}$ to the PDF set: $\alpha_{s}(M_{Z})_{\text{MSTW2008LO}} \hspace{1mm}\Rightarrow \hspace{1mm} \Lambda_{\text{QCD}} = 0.255 \hspace{1mm} \text{GeV}\hspace{1mm}$. For computations with $Q = 3$ GeV $<m_{b}\approx 4-5$ GeV, we use $n_{f} =4$ instead of $5$. 

\begin{figure}[htb]
     \centering
     \begin{subfigure}[b]{0.3\textwidth}
         \centering
         \includegraphics[width=\textwidth]{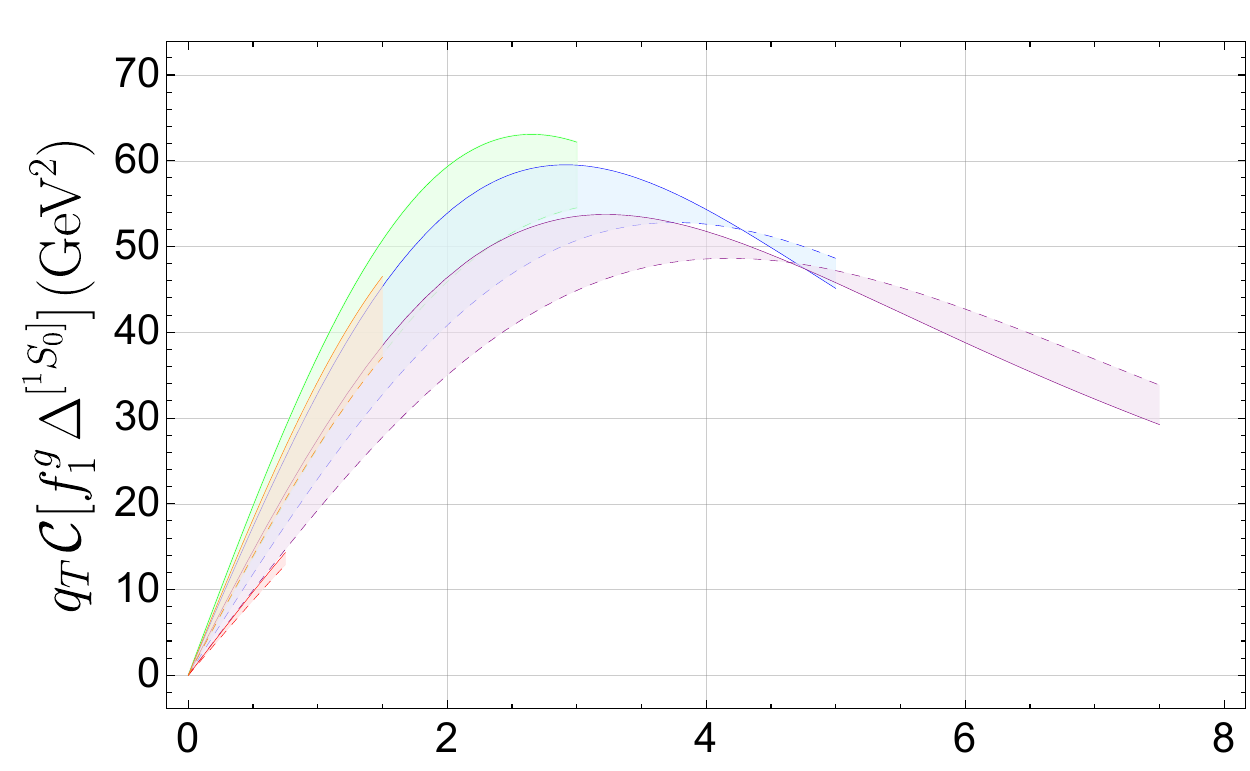}
     \end{subfigure}
     \hfill
     \begin{subfigure}[b]{0.3\textwidth}
         \centering
         \includegraphics[width=\textwidth]{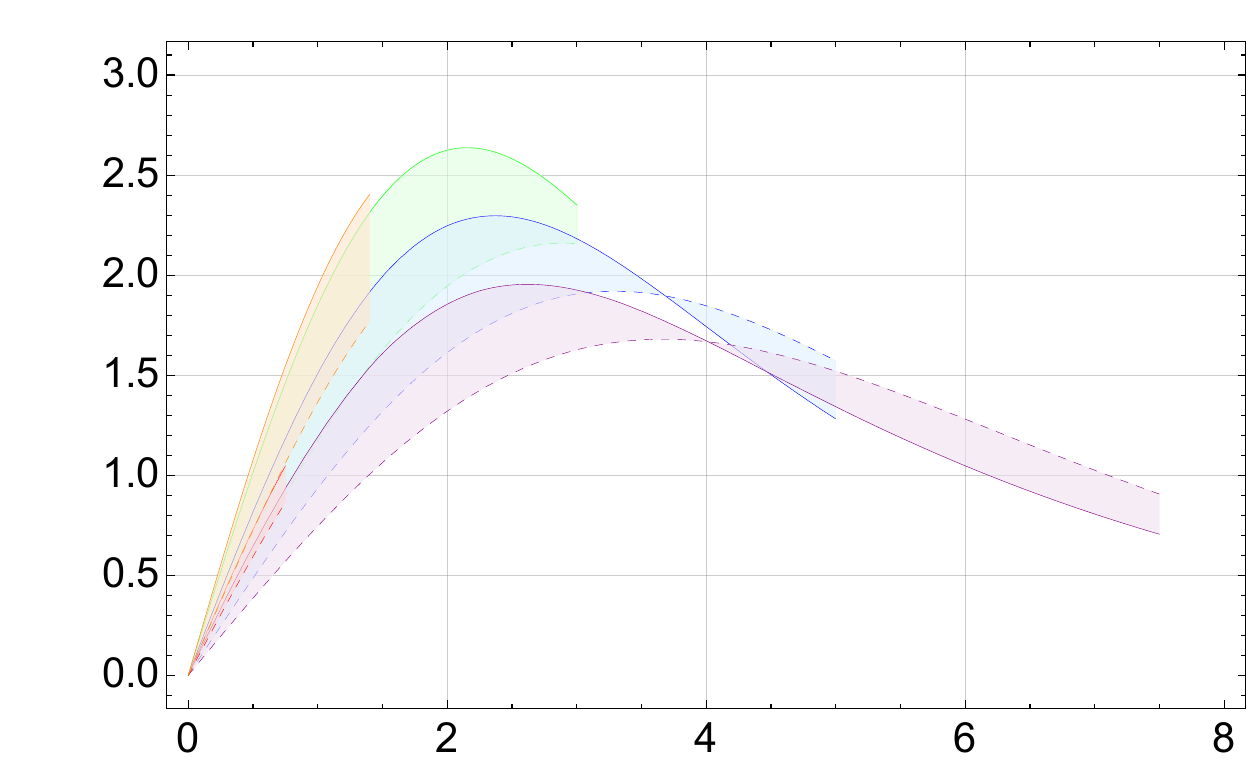}
     \end{subfigure}
     \hfill
     \begin{subfigure}[b]{0.3\textwidth}
         \centering
         \includegraphics[width=\textwidth]{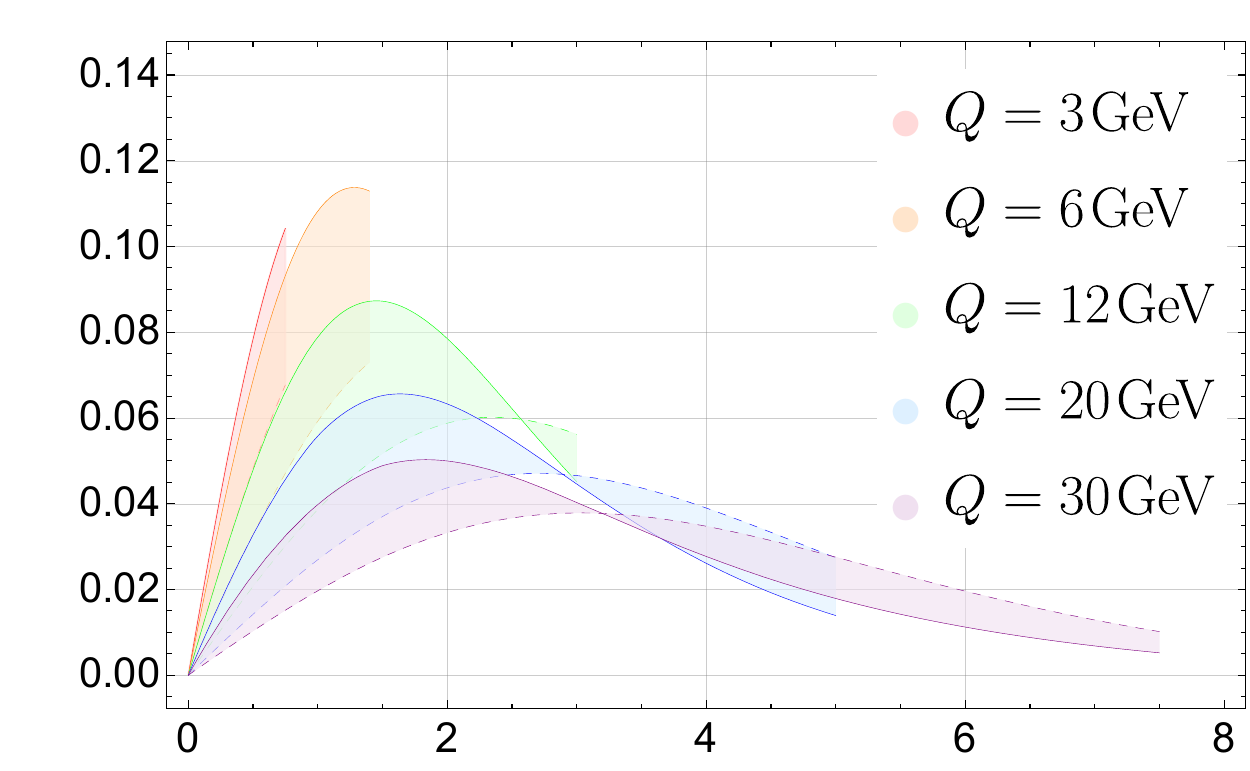}
     \end{subfigure}
          \begin{subfigure}[b]{0.3\textwidth}
         \centering
         \includegraphics[width=\textwidth]{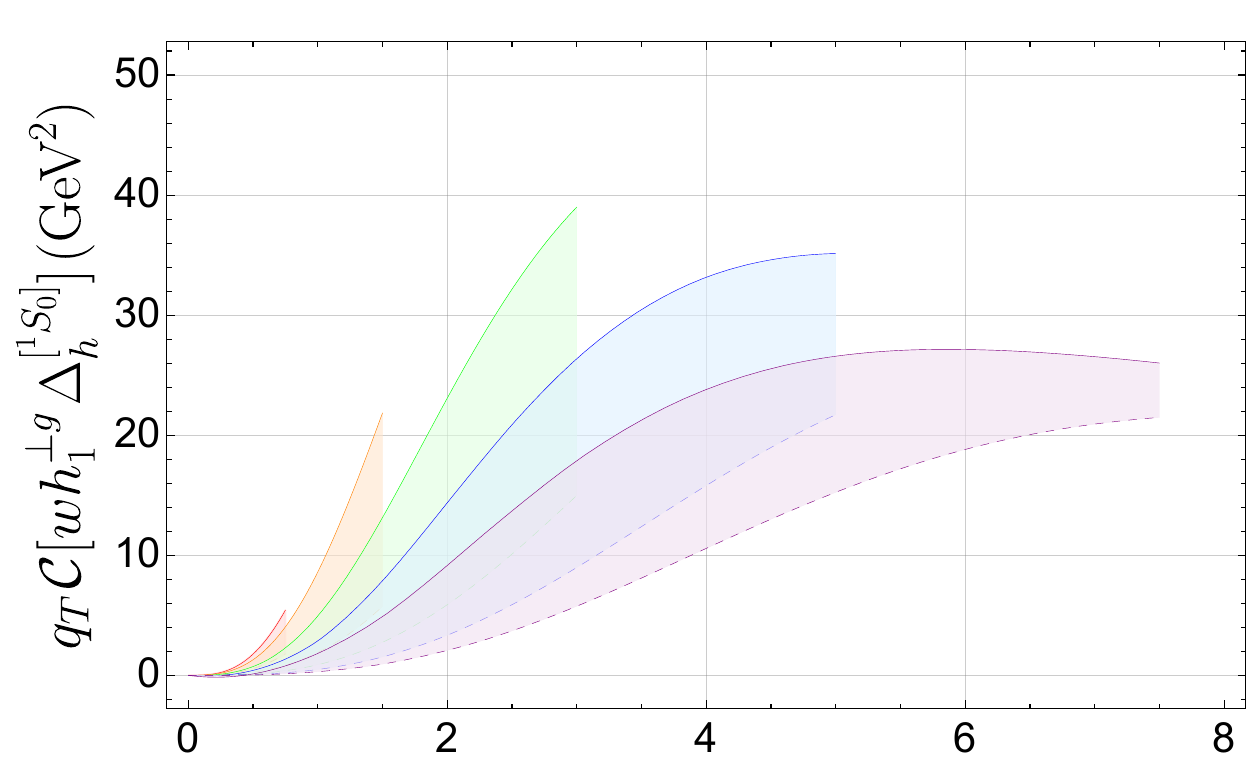}
     \end{subfigure}
     \hfill
     \begin{subfigure}[b]{0.3\textwidth}
         \centering
         \includegraphics[width=\textwidth]{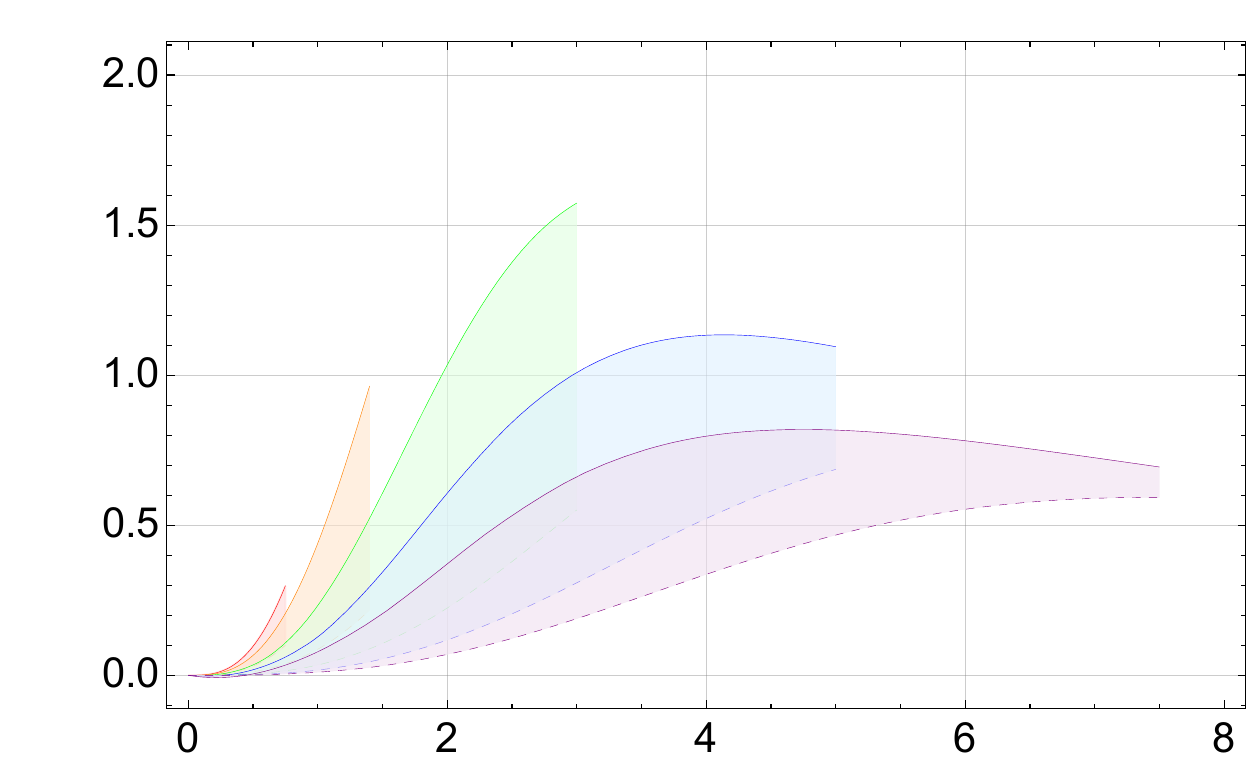}
     \end{subfigure}
     \hfill
     \begin{subfigure}[b]{0.3\textwidth}
         \centering
         \includegraphics[width=\textwidth]{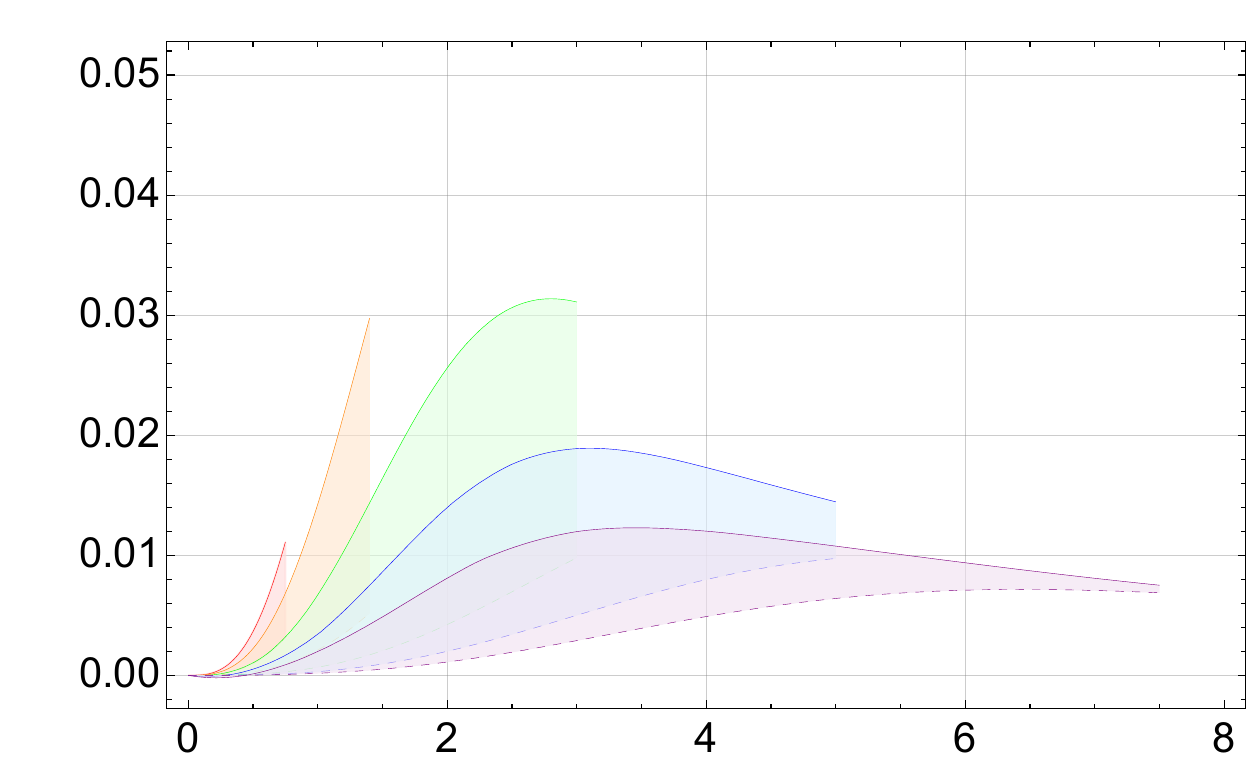}
     \end{subfigure}
          \begin{subfigure}[b]{0.3\textwidth}
         \centering
         \includegraphics[width=\textwidth]{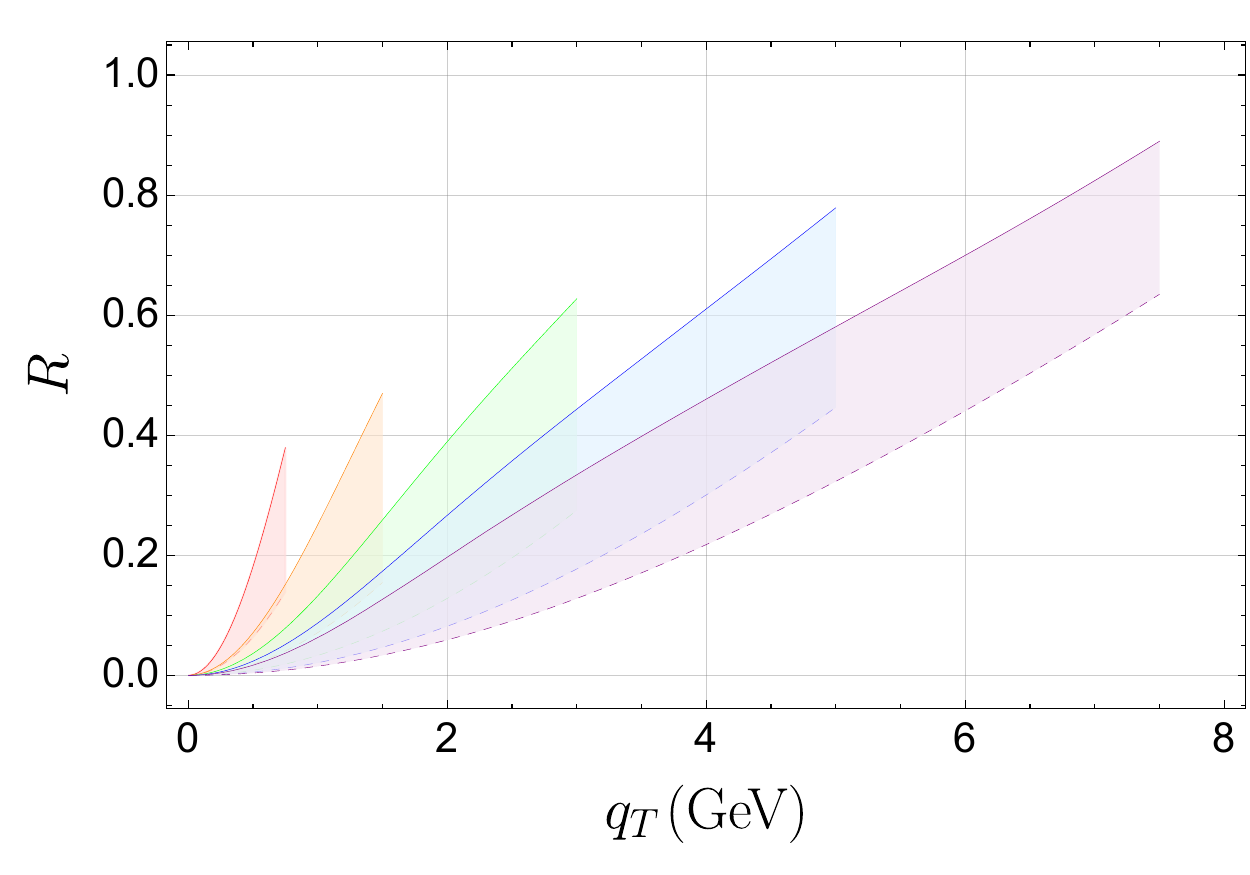}
     \end{subfigure}
     \hfill
     \begin{subfigure}[b]{0.3\textwidth}
         \centering
         \includegraphics[width=\textwidth]{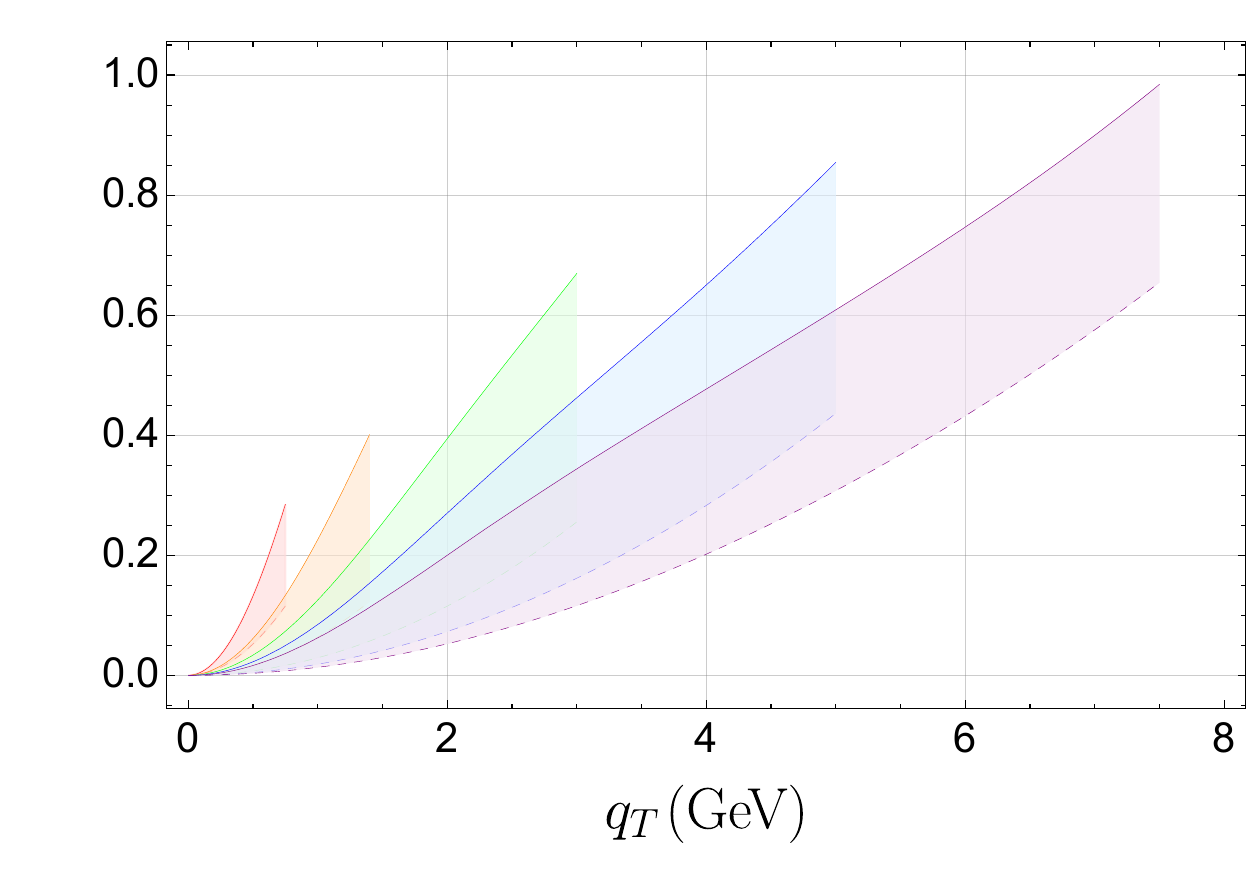}
     \end{subfigure}
     \hfill
     \begin{subfigure}[b]{0.3\textwidth}
         \centering
         \includegraphics[width=\textwidth]{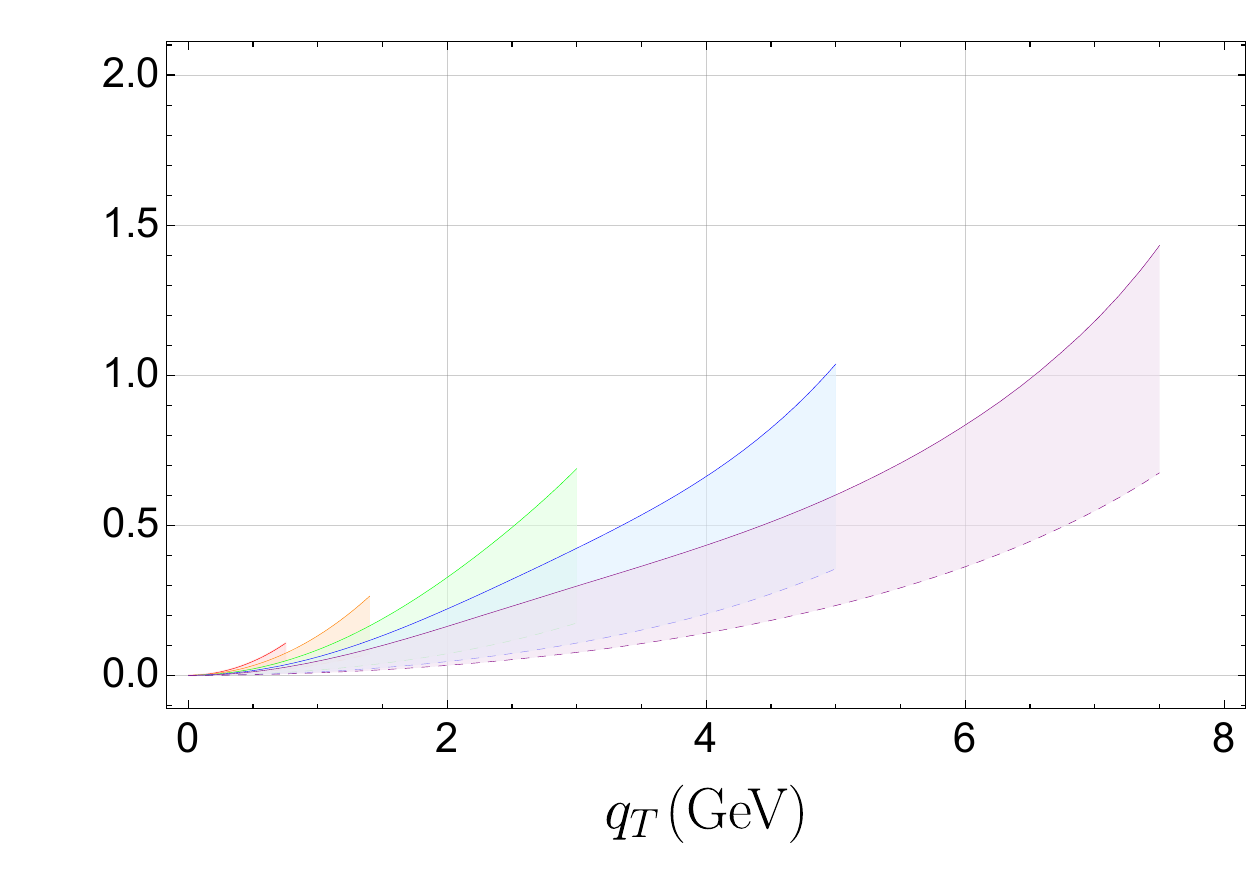}
     \end{subfigure}
    \caption{The convolutions (using $\langle 0| \OP_{8}^{J/\psi}(^{1}S_{0})|0 \rangle$ from CMSWZ) times $q_{T}$ and the ratio of the convolutions $R$ as a function of $q_{T}$ for different values of $Q$ with $x=10^{-3}$ (left), $x=10^{-2}$ (middle) and $x=10^{-1}$ (right) using $b_{T,\text{lim}}=[2:8]\hspace{1mm}\text{GeV}^{-1}$.}
\label{illustration}
\end{figure}

As an illustration of the typical features of the TMD evolution of the convolutions $q_{T}\hspace{1mm}\Co[f_{1}^{\perp g}\hspace{1mm}\Delta^{[n]}]$, $q_{T}\hspace{1mm}\Co[wh_{1}^{\perp g}\hspace{1mm}\Delta^{[n]}_{h}]$, and their ratio, in Figure \ref{illustration} we show the results for one particular LDME, namely $\langle 0| \OP_{8}^{J/\psi}(^{1}S_{0})|0 \rangle$ from CMSWZ. Results are shown for a range of $Q$ values of relevance to the EIC, $Q=3$, $6$, $12$, $20$ and $30$ GeV, and for three different values of $x$: $10^{-3}$, $10^{-2}$, and $10^{-1}$. Although $x=10^{-1}$ lies outside the gluon dominated region and we do not include the contribution from quark TMDs, we include this case for illustration purposes in order to see the results for higher $Q$ values. At the two smaller $x$ values the contribution from the collinear quark PDF to the tail of $h_{1}^{\perp g}$ in Eq.\ (\ref{h1perptail}) is non-negligible, therefore, we do include that. 

We observe that the $q_{T}$-spectrum broadens and the estimated uncertainty band from the unknown nonperturbative contributions becomes smaller with increasing $Q$, as one would expect. All curves are shown for $q_T < Q/4$ in order to ensure that the positivity bound is respected. For most curves this restriction is sufficient, except for $x=10^{-1}$ the restriction $q_{T} < Q/4$ is not enough for the higher $Q$ values. Therefore, we cut off at an even lower $q_{T}$ when making azimuthal asymmetry predictions for $x=10^{-1}$.

\begin{figure}[htb]
    \centering
    \includegraphics[width=10cm]{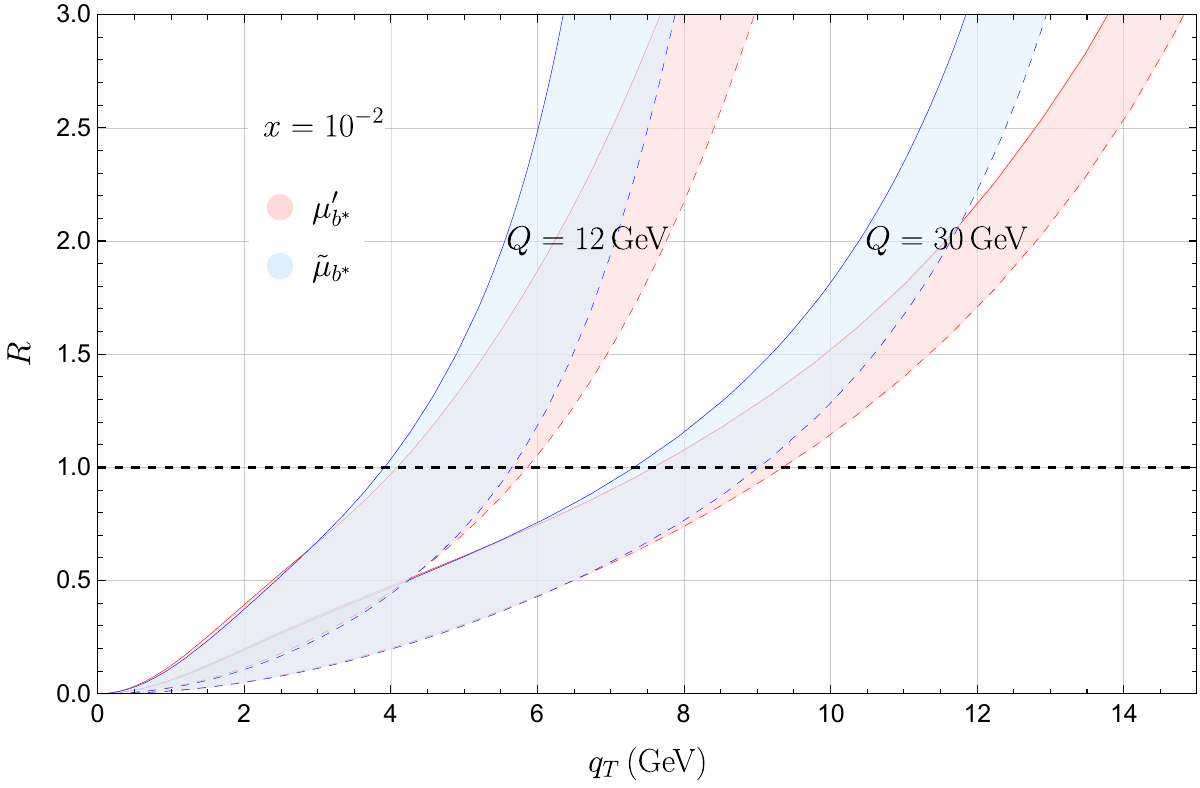}
    \caption{$R$ as a function of $q_{T}$, using $\mu_{b^*}'$ and $\tilde{\mu}_{b^*}'$. The dashed boundary line denotes the $b_{T,\text{lim}}= 2 \hspace{1mm}\text{GeV}^{-1}$ computation and the solid boundary line denotes the $b_{T,\text{lim}}= 8 \hspace{1mm}\text{GeV}^{-1}$ computation. The dashed horizontal black line denotes the positivity bound.}
\label{violation}
\end{figure}

In our computations the ratio of convolutions $R$ violates the positivity bound within what is usually expected to be the range of validity of TMD factorization $q_{T} < Q/2$, as shown in Figure \ref{violation}. This problem originates from the very small $b$ region. This is clear from Figure \ref{violation}, where we compare the results for the two ways to ensure that $b_T \geq b_0/Q$ in the perturbative Sudakov factor that we discussed earlier: $\mu_{b} \to \mu_{b}'$ or $\mu_{b} \to \tilde{\mu}_{b}'$. The region where the asymmetry starts to become sensitive to how the very small $b$ region is treated roughly corresponds to the region where the positivity bound is violated. As far as we know this is the first instance of an azimuthal asymmetry to display sensitivity to the very small $b$ region well within what is commonly considered the TMD region. Since the replacement $\mu_{b} \to \mu_{b}'$ leads to slightly smaller asymmetries, we adopt that choice in what follows, but for $q_T < Q/4$ that choice does not really matter. 

Returning to the discussion of Figure \ref{illustration}, generally the maxima of the convolutions increase towards smaller $Q$, except for the small $Q$ cases. For $q_{T}\hspace{1mm}\Co[f_{1}^{g}\hspace{1mm}\Delta]$ this can be understood from the significant relative decrease in magnitude of the perturbative $f_{1}^{g}$ tail for smaller $Q$ values in combination with the Sudakov factors, where the $B$-term is then of large influence. The decrease of $q_{T}\hspace{1mm}\Co[wh_{1}^{\perp g}\hspace{1mm}\Delta_{h}]$ is, on the other hand, predominantly due to the Sudakov factors as the perturbative $h_{1}^{\perp g}$ tail at these $x$ values stays approximately constant with varying $Q$. 

The differences in behavior of the two convolutions depends on various factors: the differences between the tails of the TMDs, the Sudakov factors and the type of Bessel function.
The product of the tail with the Sudakov factors will go to zero at large $b_{T}$ and the presence of $h_{1}^{\perp g}$ in a convolution contributes to reducing the magnitude of the integrand and to $b_{T}$-broadening. 
The $h_{1}^{\perp g}$ tail is naturally suppressed by order $\alpha_s$ in comparison to the $f_{1}^{g}$ tail. However, $\alpha_{s}(\mu_{b_*}')$ is growing with $b_{T}$, up to its upper value $\approx \alpha_{s}(b_{0}/b_{T,\text{max}})$, and the $h_{1}^{\perp g}$ tail can become larger in comparison with the $f_{1}^{g}$ tail at large $b_{T}$. This effect becomes more pronounced for smaller $x$.

The consequence of the $b_{T}$-broadening is that more damped oscillations of the $J_{0}$ Bessel function in $\Co[f_{1}^{g}\hspace{1mm}\Delta]$ occur before the integrand becomes zero. Each additional oscillation in the integrand brings the convolution closer to zero and more oscillations fit in a given $b_{T}$-range when $q_T$ increases. Therefore, $\Co[f_{1}^{g}\hspace{1mm}\Delta]$ with smaller $Q$ decreases faster. The situation is different for $\Co[wh_{1}^{\perp g}\hspace{1mm}\Delta_{h}]$ that contains the $J_{2}$ Bessel function. This function starts its damped oscillation at zero and goes up. The consequence is that the $b_{T}$-integrals benefit from unsuppressed intermediate $b_{T}$ values. This results in a peak maximum before large $b_{T}$ oscillations will bring the convolution down toward zero in a similar way as for $\Co[f_{1}^{g}\hspace{1mm}\Delta]$. This point is at smaller $q_{T}$ for smaller $Q$ taking into account the $b_{T}$-broadening of the Sudakov factors. Another crucial difference is that the envelope of $J_{2}$ tends slower toward zero than the $J_{0}$ one with increasing $b_{T}$. The consequence is that $\Co[wh_{1}^{\perp g}\hspace{1mm}\Delta_{h}]$ falls slower than $\Co[f_{1}^{g}\hspace{1mm}\Delta]$. Hence, $R$ and the azimuthal asymmetry, always grow with $q_{T}$, but more slowly for larger $Q$. In addition, as the large $b_{T}$ values in $\Co[wh_{1}^{\perp g}\hspace{1mm}\Delta_{h}]$ are less suppressed than in $\Co[f_{1}^{g}\hspace{1mm}\Delta]$, the azimuthal asymmetry and $\Co[wh_{1}^{\perp g}\hspace{1mm}\Delta_{h}]$ are more sensitive to the changes in $S_{NP}$.

Varying $x$ in the computations changes the convolutions. First, we notice that the overall magnitude of the convolutions is smaller for larger $x$, because the collinear gluon PDF is less prominent at larger $x$. On the other hand, the magnitude of $R$ becomes higher for large $Q$, but lower for small $Q$ when increasing $x$ in the range of chosen values. Secondly, the shape of the perturbative TMD tails is different, in particular the $f_{1}^{g}$ tail is broader in $b_{T}$ for larger $x$. On top of that, the $B$-term in $S_{NP}$ is smaller, as can be seen in Table \ref{table1}. Together, the broader $b_{T}$ integrands make these convolutions go faster to zero for larger $x$. This explains the behavior of the magnitude of $R$ and that the azimuthal asymmetry become less straight at $x=10^{-1}$, especially visible for larger $Q$.

\begin{figure}[htb]
     \centering
     \begin{subfigure}[b]{0.49\textwidth}
         \centering
         \includegraphics[width=\textwidth]{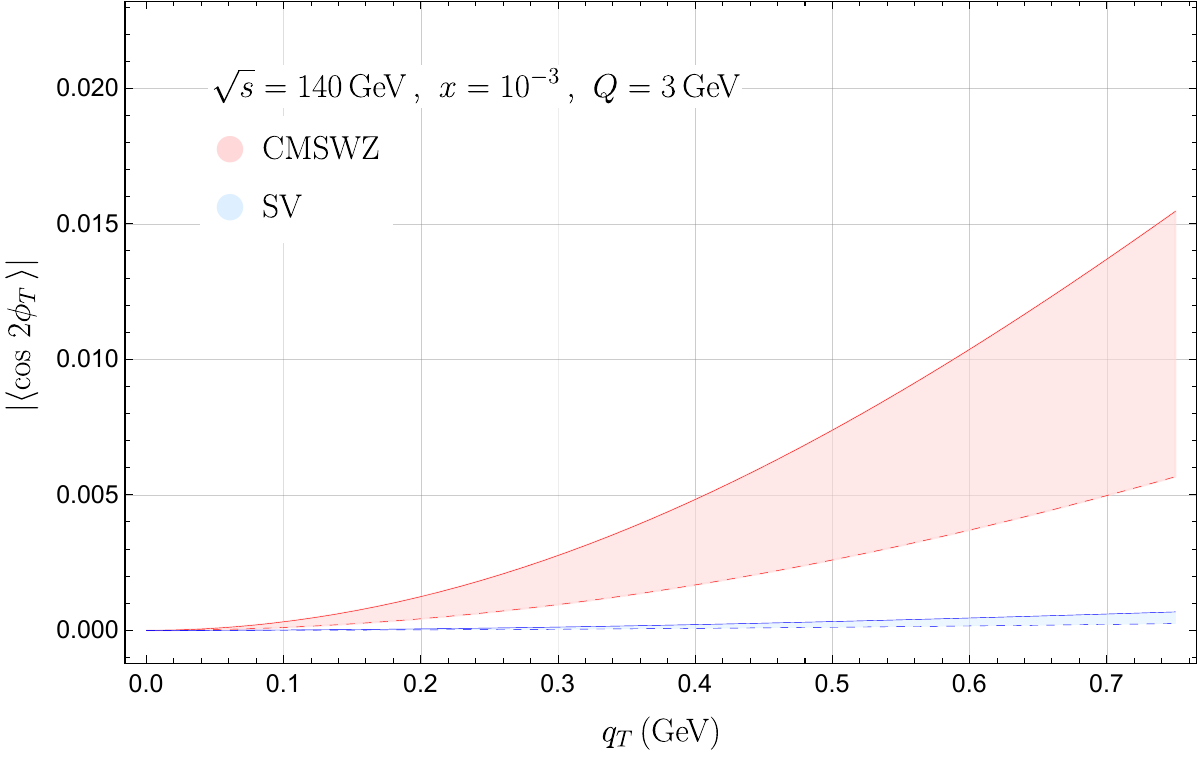}
     \end{subfigure}
     \hfill
     \begin{subfigure}[b]{0.48\textwidth}
         \centering
         \includegraphics[width=\textwidth]{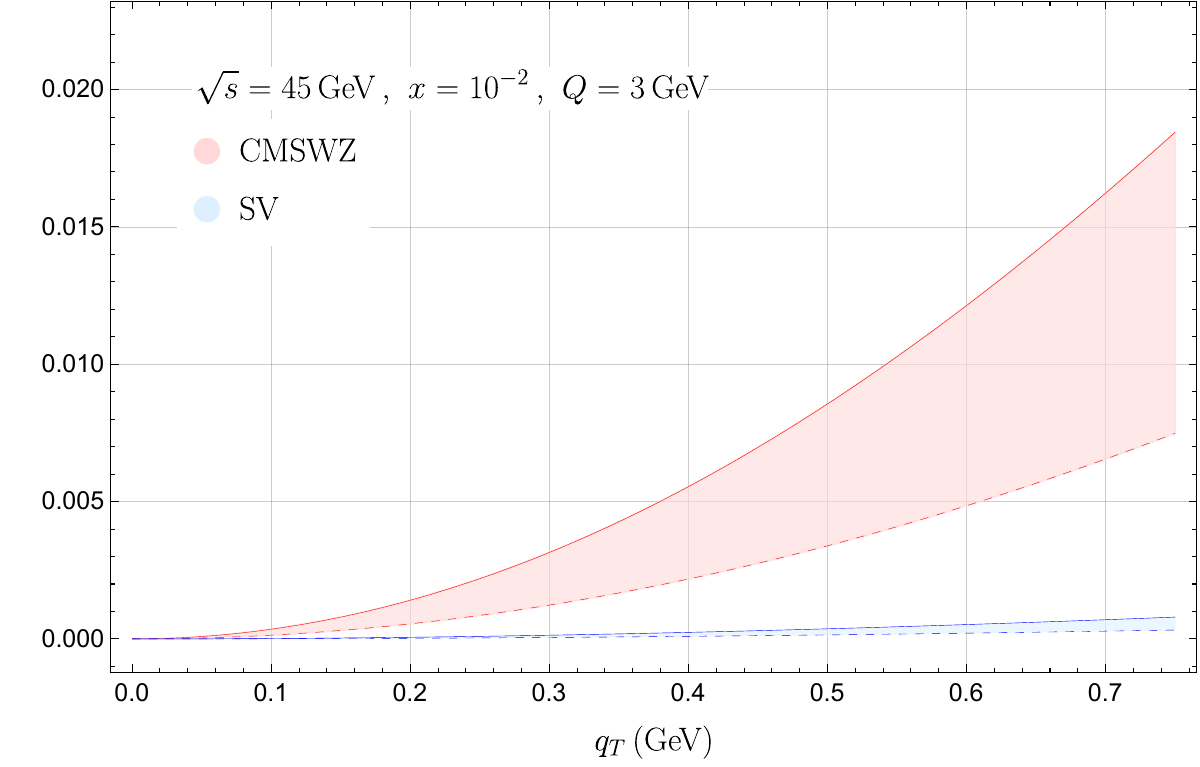}
     \end{subfigure}
        \caption{The azimuthal asymmetry as a function of $q_{T}$ for $Q=3$ GeV, and for the CMSWZ and SV LDMEs. Left is for $\sqrt{s}=140$ GeV and $x=10^{-3}$. Right is for $\sqrt{s}=45$ GeV and $x=10^{-2}$.}
        \label{LDMEeffect}
\end{figure}

\begin{figure}[htb]
     \centering
     \begin{subfigure}[b]{0.49\textwidth}
         \centering
         \includegraphics[width=\textwidth]{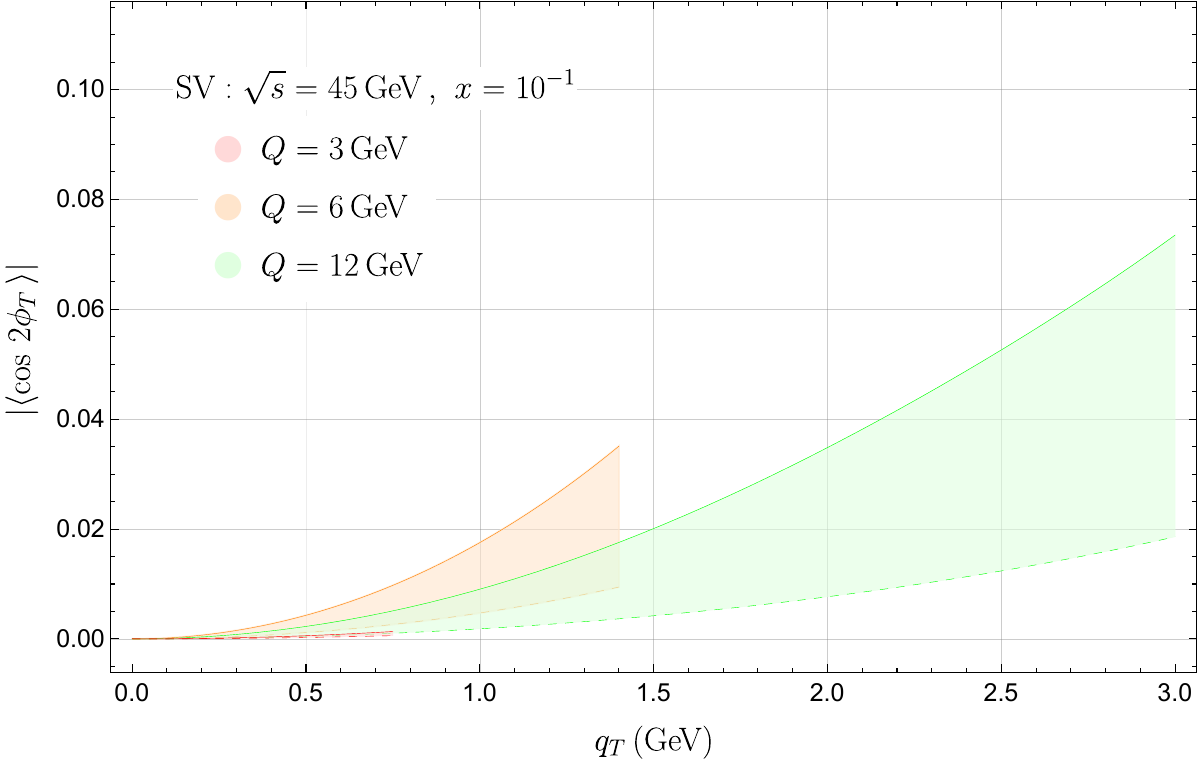}
     \end{subfigure}
     \hfill
     \begin{subfigure}[b]{0.49\textwidth}
         \centering
         \includegraphics[width=\textwidth]{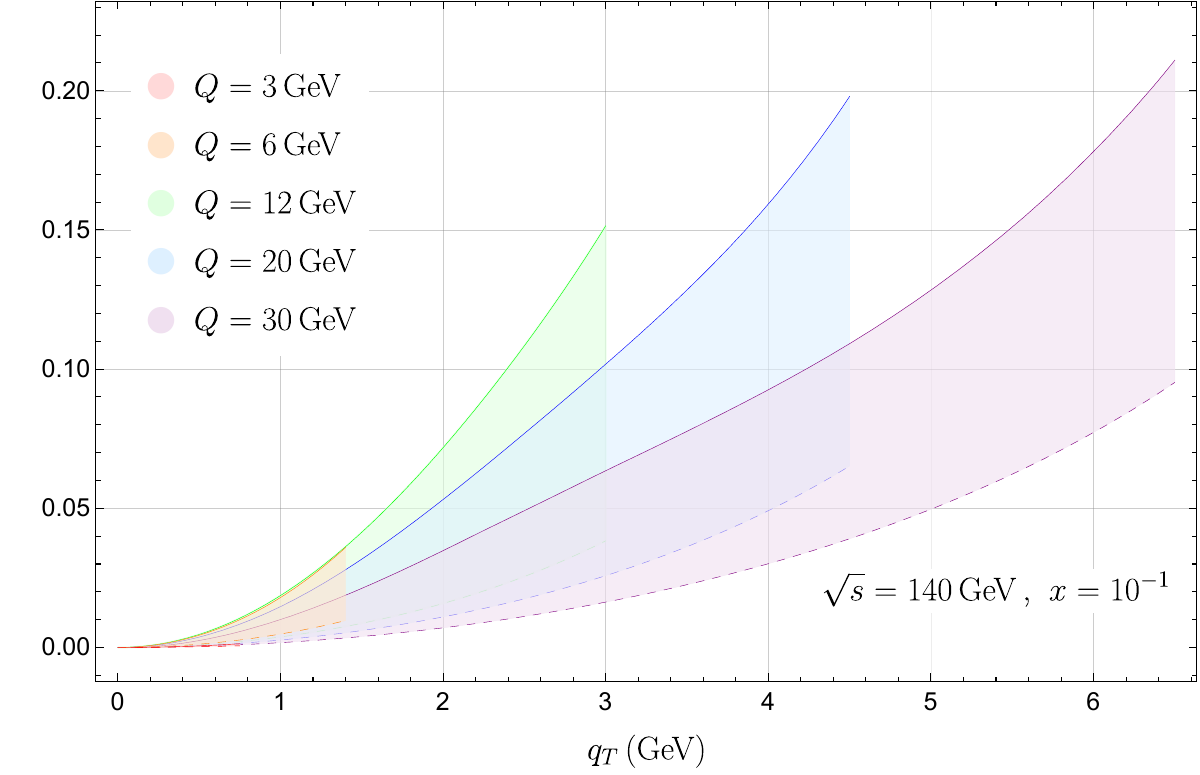}
     \end{subfigure}
        \caption{The azimuthal asymmetry as a function of $q_{T}$ for $x=10^{-1}$ using the SV LDMEs.}
        \label{EICpredictions}
\end{figure}

After these general qualitative observations, we present predictions for the azimuthal asymmetry at the EIC, shown in Figures \ref{LDMEeffect} to \ref{EICpredictions}, using the LDMEs from Table \ref{table2} and $s=Q^{2}/(x_{B}y)$ for two $\sqrt{s}$ values, $\sqrt{s}=45$ GeV and $\sqrt{s}=140$ GeV, commonly considered for the EIC. We present only results using the CMSWZ and SV LDMEs values, because $\langle 0| \OP_{8}^{J/\psi}(^{3}P_{0})|0 \rangle$ from the BK set is negative which can lead to negative cross sections for certain values of $Q$ when used in our LO expressions, and the values of BCKL are very similar to the CMSWZ values, but with larger errors. In general the asymmetry is (much) larger when using the CMSWZ set compared to the SV set, especially at small $Q$. TMD evolution predicts the azimuthal asymmetry to grow with $q_{T}$ as discussed previously, and we see that the choice of LDMEs is of large influence in the predictions. This results in an uncertainty in the predictions that is larger than that due to $S_{NP}$. In the plots we restrict to the central values of the LDME fits, but of course, taking into account the uncertainties in these values would broaden the bands further, such that the bands for CMSWZ and SV start to overlap more, giving rise to a large range of possible asymmetry values at EIC. Asymmetries anywhere between 1\% and 20\% may thus be expected at the EIC, which seem feasible to measure. Improved constraints on the LDMEs, and more generally on the TMD shape functions, can very likely be obtained in this way.  

\begin{figure}[htb]
     \centering
     \begin{subfigure}[b]{0.49\textwidth}
         \centering
         \includegraphics[width=\textwidth]{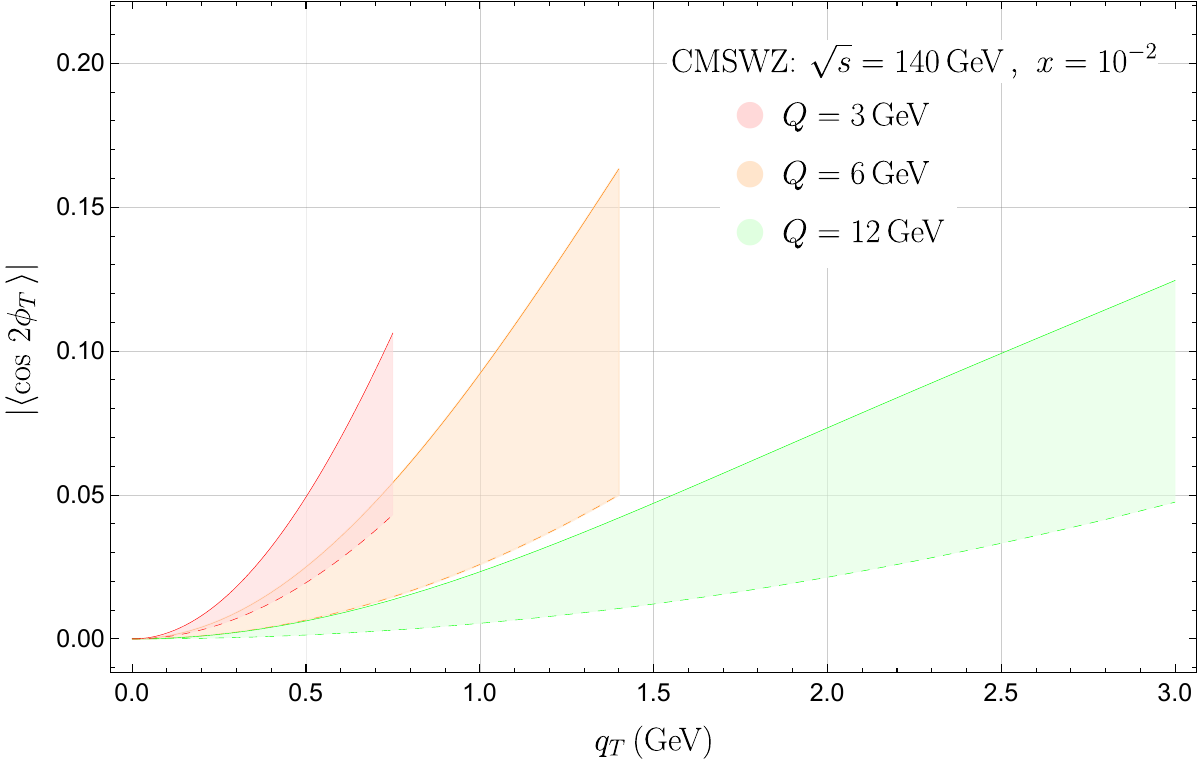}
     \end{subfigure}
     \hfill
     \begin{subfigure}[b]{0.49\textwidth}
         \centering
         \includegraphics[width=\textwidth]{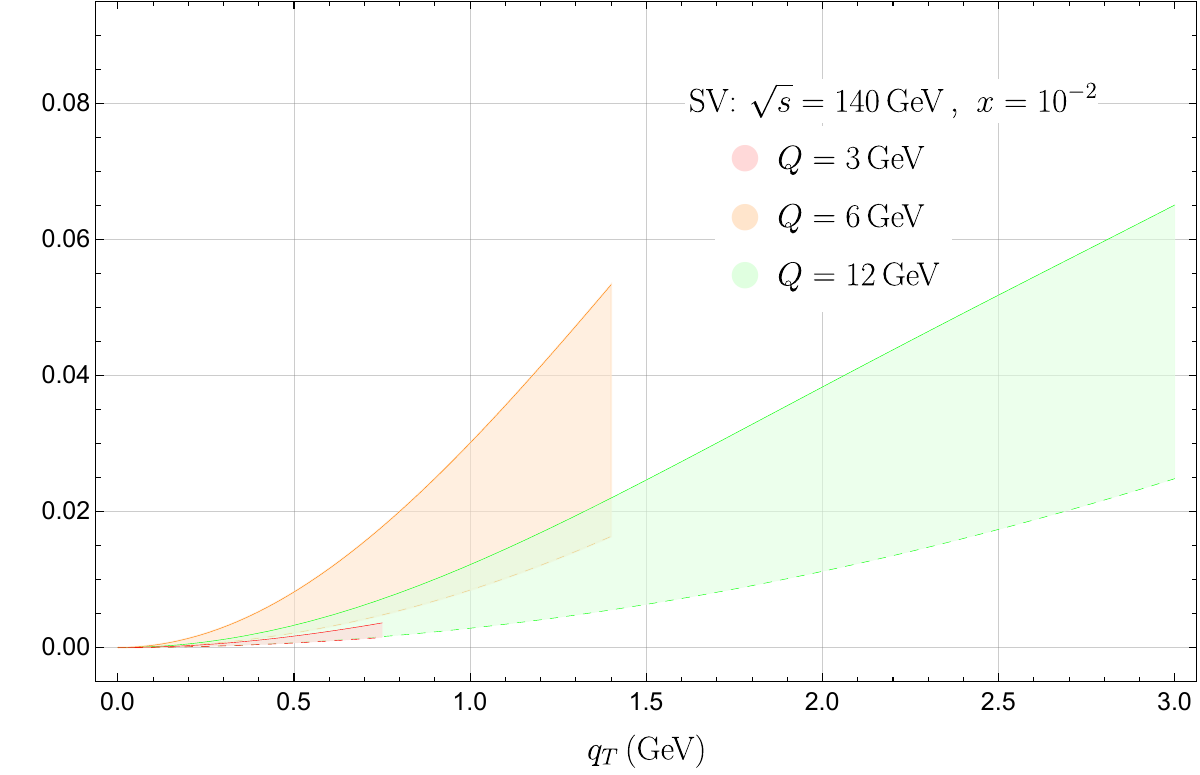}
     \end{subfigure}
        \caption{The azimuthal asymmetry as a function of $q_{T}$ for $x=10^{-2}$ using the CMSWZ (left) and SV (right) LDMEs for various $Q$ values that are kinematically allowed.}
\end{figure}

\begin{figure}[htb]
     \centering
     \begin{subfigure}[b]{0.49\textwidth}
         \centering
         \includegraphics[width=\textwidth]{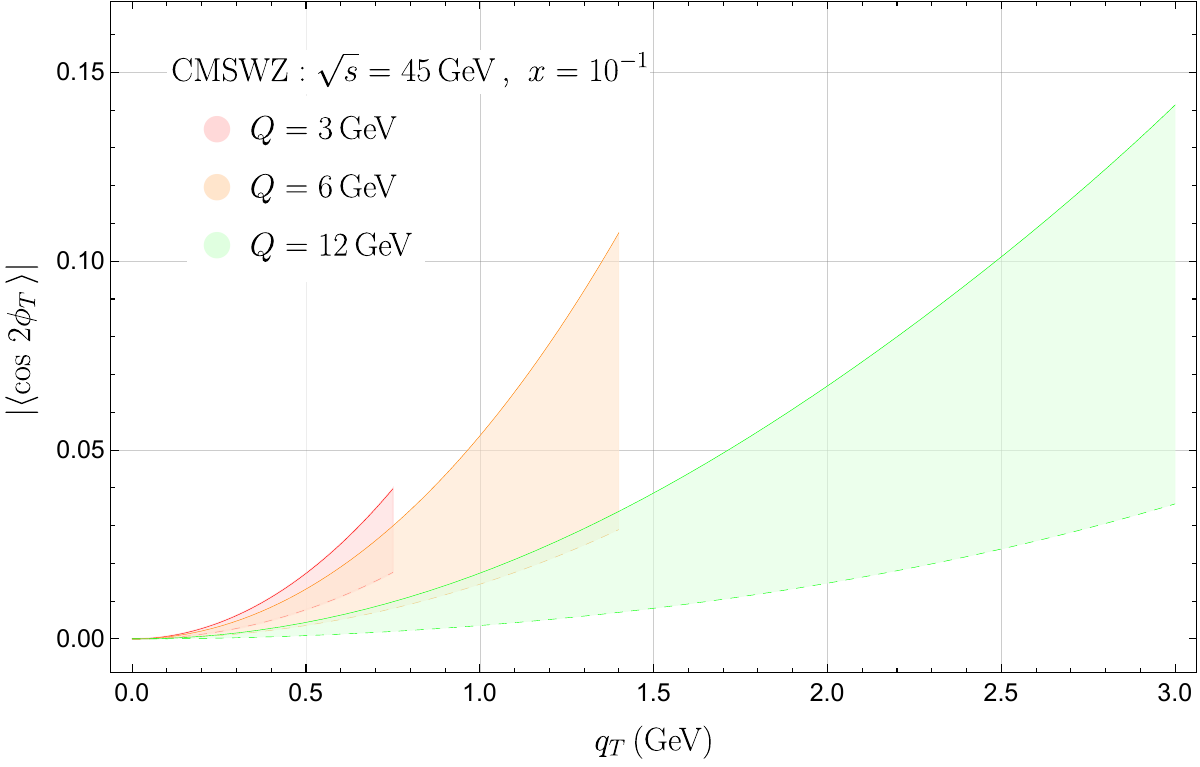}
     \end{subfigure}
     \hfill
     \begin{subfigure}[b]{0.49\textwidth}
         \centering
         \includegraphics[width=\textwidth]{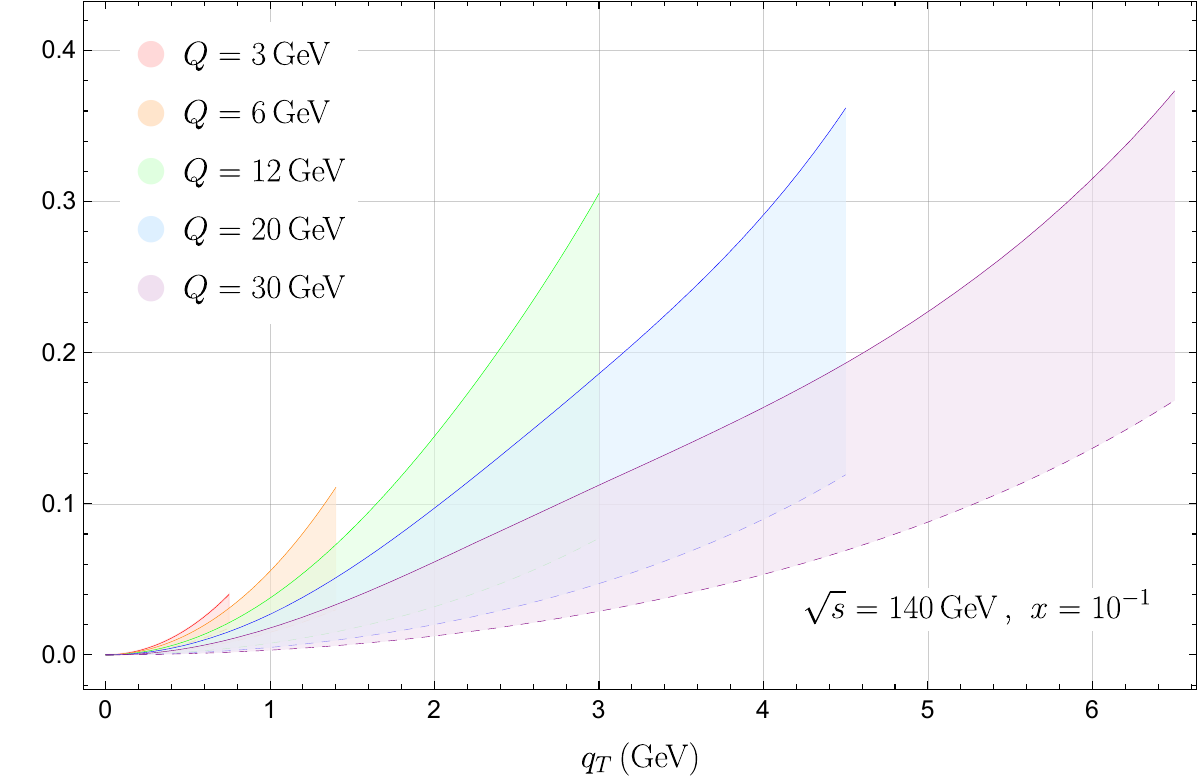}
     \end{subfigure}
        \caption{The azimuthal asymmetry as a function of $q_{T}$ for $x=10^{-1}$ using the CMSWZ LDMEs.}
\end{figure}

\newpage
\section{Discussion and conclusions}
\label{Sec5}
We have discussed the process of semi-inclusive electroproduction of $J/\psi$'s. Although the initial proton and electron beams and the final state $J/\psi$ are all unpolarized, there is an 
effect from the linear polarization of gluons inside unpolarized protons. It gives rise to a $\cos 2\phi_T$ azimuthal asymmetry for which we obtained predictions for EIC kinematics. We have included the effect of TMD evolution in order to obtain more realistic estimates. In contrast to a leading order Gaussian TMD model prediction \cite{Mukherjee:2016qxa}, we find that the azimuthal asymmetry grows monotonically in the TMD regime, more in line with the predictions from the Generalized Parton Model approach with additional gluon radiation  \cite{Kishore:2021vsm}, except that the magnitude can be much larger by as much as an order of magnitude depending on the LDME set considered. Of course, the asymmetry cannot grow beyond 1, therefore, the expectation is that a maximum will be reached outside the TMD region. We found the latter region to be significantly smaller than usually expected. For this particular process and asymmetry our computations indicate that a bound of $q_{T}< Q/4$ is required in order to respect the positivity bound and to not become sensitive to very small $b$ values in the TMD region.

Larger and monotonically rising asymmetries have also been obtained in \cite{Bacchetta:2018ivt} for the McLerran-Venugopalan model including nonlinear evolution $x$, showing decreasing asymmetries with decreasing $x$ values. In our results the $x$ dependence of the asymmetries is less systematic in the EIC kinematic range, as it depends on the considered $Q$ value and on the LDMEs. Especially SV at $Q=3$ GeV does not follow the observed trends and is exceptionally small due to a cancellation of the $S$- and $P$-wave LDMEs in Eq.\ (\ref{prefFs}) at $Q^2$ near $M^2$.

In our calculation we have restricted to leading order expressions for the small-$b_T$ expressions for the TMDs and the SFs. Assuming universality of the SFs we have adopted a perturbative Sudakov factor that was derived for $pp\to \eta_c X$ \cite{Sun:2012vc}, where there is no double logarithm associated to the SF of the heavy quarkonium state, only a single logarithm. The prefactor of the latter still needs to be confirmed by direct computation. Especially the absence of a double logarithmic contribution from the heavy quarkonium state reduces the amount of Sudakov suppression and brings the violation of the positivity bound down to smaller values of $q_T$.

We have estimated the uncertainty from the nonperturbative Sudakov factor and found that the uncertainty in the CO LDMEs forms the dominant source of uncertainty. In some cases this resulted in an order of magnitude uncertainty in the predicted asymmetry. In \cite{Boer:2021ehu} it is pointed out how one can use the polarization of the $J/\psi$ or the comparison to open heavy quark pair production to reduce this uncertainty through measurements at the EIC, but given this uncertainty, going to the next order in $\alpha_s$ in the TMD evolution calculation will not lead to much more precise predictions at the current stage. Nevertheless, the conclusion from our numerical study is that asymmetries are expected to be measurably large, especially at the larger center of mass energy of 140 GeV. 

\begin{acknowledgments}
We thank Miguel Echevarria, Jean-Philippe Lansberg, Luca Maxia, Cristian Pisano, and Feng Yuan for helpful discussions and feedback. This project has received funding from the European Union’s Horizon 2020 research and innovation programme under grant agreement No.~824093 (STRONG 2020) and is part of its JRA4-TMD-neXt Work-Package. This project has also received funding from the French Agence Nationale de la Recherche via the grant ANR-20-CE31-0015 (``PrecisOnium''). This work was also partly supported by the French CNRS via the IN2P3 project GLUE@NLO.
\end{acknowledgments}


\begin{thebibliography}{10}

\bibitem{Pisano:2013cya}
C.~Pisano, D.~Boer, S.~J. Brodsky, M.~G.~A. Buffing, and P.~J. Mulders,
  ``{Linear polarization of gluons and photons in unpolarized collider
  experiments},'' {\em JHEP}, vol.~10, p.~024, 2013.

\bibitem{Godbole:2012bx}
R.~M. Godbole, A.~Misra, A.~Mukherjee, and V.~S. Rawoot, ``{Sivers Effect and
  Transverse Single Spin Asymmetry in $e+p^\uparrow \to e+J/\psi+X$},'' {\em
  Phys. Rev.}, vol.~D85, p.~094013, 2012.

\bibitem{Godbole:2013bca}
R.~M. Godbole, A.~Misra, A.~Mukherjee, and V.~S. Rawoot, ``{Transverse Single
  Spin Asymmetry in $e+p^\uparrow \to e+J/\psi +X $ and Transverse Momentum
  Dependent Evolution of the Sivers Function},'' {\em Phys. Rev.}, vol.~D88,
  no.~1, p.~014029, 2013.

\bibitem{Godbole:2014tha}
R.~M. Godbole, A.~Kaushik, A.~Misra, and V.~S. Rawoot, ``{Transverse single
  spin asymmetry in $e+p^\uparrow \to e+J/\psi +X $ and $Q^2$ evolution of
  Sivers function-II},'' {\em Phys. Rev.}, vol.~D91, no.~1, p.~014005, 2015.

\bibitem{Zhang:2014vmh}
G.-P. Zhang, ``{Probing transverse momentum dependent gluon distribution
  functions from hadronic quarkonium pair production},'' {\em Phys. Rev.},
  vol.~D90, no.~9, p.~094011, 2014.

\bibitem{Mukherjee:2015smo}
A.~Mukherjee and S.~Rajesh, ``{Probing Transverse Momentum Dependent Parton
  Distributions in Charmonium and Bottomonium Production},'' {\em Phys. Rev.},
  vol.~D93, no.~5, p.~054018, 2016.

\bibitem{Mukherjee:2016qxa}
A.~Mukherjee and S.~Rajesh, ``{$J/\psi $ production in polarized and
  unpolarized ep collision and Sivers and $\cos 2\phi $ asymmetries},'' {\em
  Eur. Phys. J. C}, vol.~77, no.~12, p.~854, 2017.

\bibitem{DAlesio:2017rzj}
U.~D'Alesio, F.~Murgia, C.~Pisano, and P.~Taels, ``{Probing the gluon Sivers
  function in $p^\uparrow p\to J/\psi\,X$ and $p^\uparrow p \to D\,X$},'' {\em
  Phys. Rev.}, vol.~D96, no.~3, p.~036011, 2017.

\bibitem{Lansberg:2017dzg}
J.-P. Lansberg, C.~Pisano, F.~Scarpa, and M.~Schlegel, ``{Pinning down the
  linearly-polarised gluons inside unpolarised protons using quarkonium-pair
  production at the LHC},'' {\em Phys. Lett.}, vol.~B784, pp.~217--222, 2018.
\newblock [Erratum: Phys. Lett.B791,420(2019)].

\bibitem{Bacchetta:2018ivt}
A.~Bacchetta, D.~Boer, C.~Pisano, and P.~Taels, ``{Gluon TMDs and NRQCD matrix
  elements in $J/\psi$ production at an EIC},'' {\em Eur. Phys. J. C}, vol.~80,
  no.~1, p.~72, 2020.

\bibitem{Kishore:2018ugo}
R.~Kishore and A.~Mukherjee, ``{Accessing linearly polarized gluon distribution
  in $J/\psi$ production at the electron-ion collider},'' {\em Phys. Rev. D},
  vol.~99, no.~5, p.~054012, 2019.

\bibitem{Scarpa:2019fol}
F.~Scarpa, D.~Boer, M.~G. Echevarria, J.-P. Lansberg, C.~Pisano, and
  M.~Schlegel, ``{Studies of gluon TMDs and their evolution using
  quarkonium-pair production at the LHC},'' {\em Eur. Phys. J. C}, vol.~80,
  no.~2, p.~87, 2020.

\bibitem{Kishore:2022ddb}
R.~Kishore, A.~Mukherjee, A.~Powar, and M.~Siddiqah, ``{$\cos2\phi_t$ azimuthal
  asymmetry in back-to-back $J/\psi$-jet in $e~p\rightarrow e~J/\psi~Jet~ X$ at
  the EIC},'' 3 2022.

\bibitem{Mulders:2000sh}
P.~J. Mulders and J.~Rodrigues, ``{Transverse momentum dependence in gluon
  distribution and fragmentation functions},'' {\em Phys. Rev. D}, vol.~63,
  p.~094021, 2001.

\bibitem{Boer:2010zf}
D.~Boer, S.~J. Brodsky, P.~J. Mulders, and C.~Pisano, ``{Direct Probes of
  Linearly Polarized Gluons inside Unpolarized Hadrons},'' {\em Phys. Rev.
  Lett.}, vol.~106, p.~132001, 2011.

\bibitem{Boer:2016fqd}
D.~Boer, P.~J. Mulders, C.~Pisano, and J.~Zhou, ``{Asymmetries in Heavy Quark
  Pair and Dijet Production at an EIC},'' {\em JHEP}, vol.~08, p.~001, 2016.

\bibitem{STAR:2019wlg}
J.~Adam {\em et~al.}, ``{Measurement of $e^+e^-$ Momentum and Angular
  Distributions from Linearly Polarized Photon Collisions},'' {\em Phys. Rev.
  Lett.}, vol.~127, no.~5, p.~052302, 2021.

\bibitem{Boer:2012bt}
D.~Boer and C.~Pisano, ``{Polarized gluon studies with charmonium and
  bottomonium at LHCb and AFTER},'' {\em Phys. Rev. D}, vol.~86, p.~094007,
  2012.

\bibitem{Collins:2011zzd}
J.~C. Collins, {\em {Foundations of perturbative QCD}}, vol.~32.
\newblock Cambridge University Press, 11 2013.

\bibitem{Echevarria:2011epo}
M.~G. Echevarria, A.~Idilbi, and I.~Scimemi, ``{Factorization Theorem For
  Drell-Yan At Low $q_T$ And Transverse Momentum Distributions
  On-The-Light-Cone},'' {\em JHEP}, vol.~07, p.~002, 2012.

\bibitem{Echevarria:2012js}
M.~G. Echevarria, A.~Idilbi, and I.~Scimemi, ``{Soft and Collinear
  Factorization and Transverse Momentum Dependent Parton Distribution
  Functions},'' {\em Phys. Lett. B}, vol.~726, pp.~795--801, 2013.

\bibitem{Boer:2014tka}
D.~Boer and W.~J. den Dunnen, ``{TMD evolution and the Higgs transverse
  momentum distribution},'' {\em Nucl. Phys. B}, vol.~886, pp.~421--435, 2014.

\bibitem{Boer:2020bbd}
D.~Boer, U.~D'Alesio, F.~Murgia, C.~Pisano, and P.~Taels, ``{$J/\psi$ meson
  production in SIDIS: matching high and low transverse momentum},'' {\em
  JHEP}, vol.~09, p.~040, 2020.

\bibitem{Sun:2012vc}
P.~Sun, C.~P. Yuan, and F.~Yuan, ``{Heavy Quarkonium Production at Low Pt in
  NRQCD with Soft Gluon Resummation},'' {\em Phys. Rev. D}, vol.~88, p.~054008,
  2013.

\bibitem{Zhu:2013yxa}
R.~Zhu, P.~Sun, and F.~Yuan, ``{Low Transverse Momentum Heavy Quark Pair
  Production to Probe Gluon Tomography},'' {\em Phys. Lett. B}, vol.~727,
  pp.~474--479, 2013.

\bibitem{Bodwin:1994jh}
G.~T. Bodwin, E.~Braaten, and G.~P. Lepage, ``{Rigorous QCD analysis of
  inclusive annihilation and production of heavy quarkonium},'' {\em Phys. Rev.
  D}, vol.~51, pp.~1125--1171, 1995.
\newblock [Erratum: Phys.Rev.D 55, 5853 (1997)].

\bibitem{Echevarria:2019ynx}
M.~G. Echevarria, ``{Proper TMD factorization for quarkonia production:
  $pp\to\eta_{c,b}$ as a study case},'' {\em JHEP}, vol.~10, p.~144, 2019.

\bibitem{Fleming:2019pzj}
S.~Fleming, Y.~Makris, and T.~Mehen, ``{An effective field theory approach to
  quarkonium at small transverse momentum},'' {\em JHEP}, vol.~04, p.~122,
  2020.

\bibitem{Bodwin:2005hm}
G.~T. Bodwin, E.~Braaten, and J.~Lee, ``{Comparison of the color-evaporation
  model and the NRQCD factorization approach in charmonium production},'' {\em
  Phys. Rev. D}, vol.~72, p.~014004, 2005.

\bibitem{Fleming:1997fq}
S.~Fleming and T.~Mehen, ``{Leptoproduction of $J/\psi$},'' {\em Phys. Rev. D},
  vol.~57, pp.~1846--1857, 1998.

\bibitem{Brodkorb:1994de}
T.~Brodkorb and E.~Mirkes, ``{Complete $\OP(\alpha_{s}^{2})$ corrections to $(2
  + 1)$ jet cross sections in deep inelastic scattering},'' {\em Z. Phys. C},
  vol.~66, pp.~141--150, 1995.

\bibitem{Collins:1981uk}
J.~C. Collins and D.~E. Soper, ``{Back-To-Back Jets in QCD},'' {\em Nucl. Phys.
  B}, vol.~193, p.~381, 1981.
\newblock [Erratum: Nucl.Phys.B 213, 545 (1983)].

\bibitem{delCastillo:2021znl}
R.~F. del Castillo, M.~G. Echevarria, Y.~Makris, and I.~Scimemi, ``{Transverse
  momentum dependent distributions in dijet and heavy hadron pair production at
  EIC},'' {\em JHEP}, vol.~03, p.~047, 2022.

\bibitem{Aybat:2011zv}
S.~M. Aybat and T.~C. Rogers, ``{TMD Parton Distribution and Fragmentation
  Functions with QCD Evolution},'' {\em Phys. Rev. D}, vol.~83, p.~114042,
  2011.

\bibitem{Echevarria:2022}
M.~G. Echevarria {\em Talk presented at Transversity 2022, Pavia, May 23-27},
  2022.

\bibitem{Collins:2016hqq}
J.~Collins, L.~Gamberg, A.~Prokudin, T.~C. Rogers, N.~Sato, and B.~Wang,
  ``{Relating Transverse Momentum Dependent and Collinear Factorization
  Theorems in a Generalized Formalism},'' {\em Phys. Rev. D}, vol.~94, no.~3,
  p.~034014, 2016.

\bibitem{Collins:1984kg}
J.~C. Collins, D.~E. Soper, and G.~F. Sterman, ``{Transverse Momentum
  Distribution in Drell-Yan Pair and W and Z Boson Production},'' {\em Nucl.
  Phys. B}, vol.~250, pp.~199--224, 1985.

\bibitem{Sun:2011iw}
P.~Sun, B.-W. Xiao, and F.~Yuan, ``{Gluon Distribution Functions and Higgs
  Boson Production at Moderate Transverse Momentum},'' {\em Phys. Rev. D},
  vol.~84, p.~094005, 2011.

\bibitem{Chao:2012iv}
K.-T. Chao, Y.-Q. Ma, H.-S. Shao, K.~Wang, and Y.-J. Zhang, ``{$J/\psi$
  Polarization at Hadron Colliders in Nonrelativistic QCD},'' {\em Phys. Rev.
  Lett.}, vol.~108, p.~242004, 2012.

\bibitem{Sharma:2012dy}
R.~Sharma and I.~Vitev, ``{High transverse momentum quarkonium production and
  dissociation in heavy ion collisions},'' {\em Phys. Rev. C}, vol.~87, no.~4,
  p.~044905, 2013.

\bibitem{Butenschoen:2010rq}
M.~Butenschoen and B.~A. Kniehl, ``{Reconciling $J/\psi$ production at HERA,
  RHIC, Tevatron, and LHC with NRQCD factorization at next-to-leading order},''
  {\em Phys. Rev. Lett.}, vol.~106, p.~022003, 2011.

\bibitem{Bodwin:2014gia}
G.~T. Bodwin, H.~S. Chung, U.-R. Kim, and J.~Lee, ``{Fragmentation
  contributions to $J/\psi$ production at the Tevatron and the LHC},'' {\em
  Phys. Rev. Lett.}, vol.~113, no.~2, p.~022001, 2014.

\bibitem{Martin:2009iq}
A.~D. Martin, W.~J. Stirling, R.~S. Thorne, and G.~Watt, ``{Parton
  distributions for the LHC},'' {\em Eur. Phys. J. C}, vol.~63, pp.~189--285,
  2009.

\bibitem{Kishore:2021vsm}
R.~Kishore, A.~Mukherjee, and M.~Siddiqah,
  ``{$\text{Cos}\hspace{1mm}(2\phi_{h})$ asymmetry in $J/\psi$ production in
  unpolarized $ep$ collision},'' {\em Phys. Rev. D}, vol.~104, no.~9,
  p.~094015, 2021.

\bibitem{Boer:2021ehu}
D.~Boer, C.~Pisano, and P.~Taels, ``{Extracting color octet NRQCD matrix
  elements from $J/\psi$ production at the EIC},'' {\em Phys. Rev. D},
  vol.~103, no.~7, p.~074012, 2021.

\end{thebibliography}
\end{document}